\documentclass[a4paper,11pt]{article}

\usepackage{amsfonts,amsmath,amssymb}

\def\boxit#1{\vbox{\hrule\hbox{\vrule\kern3pt
    \vbox{\kern3pt#1\kern3pt}\kern3pt\vrule}\hrule}}

\usepackage{graphicx}
\usepackage[T1]{fontenc}

\newtheorem{prop}{Proposition}
\newtheorem{coro}{Corollary}
\newtheorem{lem}{Lemma}

\newtheorem{theo}{Theorem}

\def\fdm{\hfill$\bbox$}
\def\is{\textsc{max independent set}}
\def\3is{\textsc{max independent set-3}}
\def\4is{\textsc{max independent set-4}}
\def\5is{\textsc{max independent set-5}}
\def\6is{\textsc{max independent set-6}}

\def\pr{\noindent \textbf{Proof. }}

\newcommand{\bbox}{\vrule height7pt width4pt depth1pt}
\newcommand{\runningtime}{1.08537}

\title{\textbf{Fast algorithms for \is{} in graphs of small average degree}}

\author{N. Bourgeois${}^1$ \hspace*{1cm} B. Escoffier${}^1$ \hspace*{1cm} V.~Th.~Paschos${}^1$ \hspace*{1cm} J.M.M. van Rooij${}^2$ \vspace*{4mm} \\
${}^1$~LAMSADE, CNRS UMR~7024 and Universit\'e Paris-Dauphine, France \\
\texttt{\{bourgeois,escoffier,paschos\}@lamsade.dauphine.fr} \vspace*{2mm} \\
${}^2$~Department of Information and Computing Sciences \\
Universiteit Utrecht, The Netherlands \\
\texttt{johanvr@cs.uu.nl}}

\begin{document}

\maketitle

\begin{abstract}

\is{} is a paradigmatic problem in theoretical computer science and
numerous studies tackle its resolution by exact algorithms with
non-trivial worst-case complexity. The best such complexity is, to
our knowledge, the~$O^*(1.1889^n)$ algorithm claimed by (J.~M.
Robson, \textit{Finding a maximum independent set in
time~${O}(2^{n/4})$}, Technical Report 1251-01, LaBRI, Université de
Bordeaux~I, 2001) in his unpublished technical report. We also quote
the~$O^*(1.2210^n)$ algorithm by ({F.~V.}~Fomin, F.~Grandoni and
D.~Kratsch, \textit{Measure and conquer: a simple~${O}(2^{0.288n})$
independent set algorithm}, Proc. SODA'06, pages 18--25, 2006), that
is the best published result about \is{}.
In this paper we settle \is{} in (connected) graphs with ``small'' average degree, more precisely with average degree at most~3, 4, 5 and~6. 
Dealing with exact computation of \is{} in  graphs of average degree
at most 3, the best bound known is the recent~$O^*(1.0977^n)$ bound
by (N.~Bourgeois, B.~Escoffier and V.~Th. Paschos,
\textit{An~${O}^*(1.0977^n)$ exact algorithm for \textsc{max
independent set} in sparse graphs}, Proc. IWPEC'08, LNCS~5018, pages
55--65, 2008). Here we improve this result down to~$O^*(1.0854^n)$
by proposing finer and more powerful reduction rules.
We then propose a generic method showing how improvement of the
worst-case complexity for \is{} in graphs of average degree~$d$
entails improvement of it in any graph of average degree greater
than~$d$ and, based upon it, we tackle \is{} in graphs of average
degree 4, 5 and~6.
For \is{} in graphs with average degree~4, we provide an upper
complexity bound of~$O^*(1.1571^n)$, obviously still valid for
graphs of maximum degree 4, that outperforms the best known bound
of~$O^*(1.1713^n)$ by~(R.~Beigel, \textit{Finding maximum
independent sets in sparse and general graphs}, Proc. SODA'99, pages
856--857, 1999).
For \is{} in graphs of average degree at most 5 and 6, we provide
bounds of~$O^*(1.1969^n)$ and~$O^*(1.2149^n)$, respectively, that
improve upon the corresponding bounds of~$O^*(1.2023^n)$
and~$O^*(1.2172^n)$ in graphs of maximum degree 5 and 6 by (Fomin et
al.,~2006).
Let us remark that in the cases of graphs of average degree at most
3 and~4, our bounds outerperform the~$O^*(1.1889^n)$ claimed by
(Robson,~2001).

\end{abstract}

\section{Introduction}

Very active research has been recently conducted around the
development of optimal algorithms for \textbf{NP}-hard problems with
non-trivial worst-case complexity (see the seminal paper
by~\cite{Woeg2} for a survey on both methods used and results
obtained). Among the problems studied in this field, \is{} (and
particular versions of it) is one of those that have received a very
particular attention and made much effort spent.

Given a graph $G = (V,E)$, \is{} consists of finding a
max\-i\-mum-si\-ze subset $V' \subseteq V$ such that for any
$(v_i,v_j) \in V' \times V'$, $(v_i,v_j) \notin E$. For this problem
the best worst-case complexity bound is, to our knowledge,
the~$O^*(1.1889^n)$ bound claimed by~\cite{robson01} in his
unpublished technical report. We also quote the~$O^*(1.2210^n)$
algorithm by~\cite{fomin_et_al_soda_06}, that is the best published
result about \is{}.

Let~$T(\cdot)$ be a super-polynomial and~$p(\cdot)$ be a polynomial, both on integers. In what follows, using notations in~\cite{Woeg2}, for an integer~$n$, we express running-time bounds of the form~$p(n)\cdot T(n)$ as~$O^*(T(n))$, the star meaning that we ignore polynomial factors. We denote by~$T(n)$ the worst-case time required to exactly solve the considered combinatorial optimization problem on an instance of size~$n$. We recall (see, for instance,~\cite{eppsoda01}) that, if it is possible to bound above~$T(n)$ by a recurrence expression of the type~$T(n) \leq \sum T(n-r_i) + O(p(n))$, we have~$\sum T(n-r_i) +  O(p(n)) = O^*(\alpha(r_1,r_2,\ldots)^n)$ where~$\alpha(r_1,r_2,\ldots)$ is the largest root of the function~$f(x) = 1 - \sum x^{-r_i}$.

In this paper we settle \is{} in (connected) graphs with ``small'' average degree, more precisely with average degree at most~3, 4, 5 and~6. Let us denote by \3is{}, -4, -5 and~-6, the restrictions of \is{} to graphs of maximum degree~3, 4, 5 and~6, respectively.

For \3is{}, several algorithms have been devised, successively improving its worst case complexity. Let us quote the~$O^*(1.1259^n)$ algorithm by~\cite{beigelsoda}, the~$O^*(1.1254)$ algorithm by~\cite{chekaxiissac}, the~$O^*(1.1120)$ algorithm by~\cite{fu3is}, the~$O^*(1.1034^n)$ algorithm by~\cite{Razgon06} and, finally, the recent~$O^*(1.0977^n)$ algorithm by~\cite{3-is}. As a first result, in this article we improve the bound of~\cite{3-is} down to~$O^*(1.0854^n)$ by proposing finer and more powerful reduction rules (Section~\ref{secdeg3}). Our result remains valid also for graphs of average degree bounded by~3.

We then propose a generic method extending improvements of the
worst-case complexity for \is{} in graphs of average degree~$d$ to
graphs of average degree greater than~$d$. This ``bottom-up'' method
of carrying improvements of time-bounds for restrictive cases of a
problem to less restrictive ones (the latter including the former)
is, as far as we know, a new method that can be very useful for
strengthening time-bounds not only for \is{} but also for other
graph-problems where local worst configurations appear when maximum degree
is small. For instance, when tackling \is{} in graphs of maximum degree,
say, at least~10, a simple tree-search based algorithm
with a branching rule of the form either don't take a vertex of
degree~10, or take it and remove it as well as its neighbors
(in this case~11 vertices are removed in total) guarantees an upper time-bound of~$O^*(1.1842^n)$
dominating so the bound by~\cite{robson01}. 

In order to informally sketch the method, suppose that one knows how
to solve the problem on graphs with average degree~$d$ in
time~$O^*(\gamma_d^n)$. Solving the problem on graphs with average
degree $d'\geq d$ is based upon two ideas: we first look for
complexity expression of the form~$\alpha^m\beta^n$, where~$\alpha$
and~$\beta$ depend both on the input graph, (namely on its average
degree) \emph{and on the value~$\gamma_d$} (see for instance
Section~\ref{secdeg4}). In other words, the form of the complexity
we look for is parameterized by what we already know on graphs with
smaller average degrees. Next, according to this form, we identify
particular values~$d_i$ (not necessarily integer, see for instance
Section~\ref{secdeg5}) of the average degree that ensure that a
``good'' branching occurs. This allows to find a good complexity for
increasing values of the average degree. Note also that a particular
interest of this method lies in the fact that any improvement on the
worst-case complexity on graphs of average degree~3 immediately
yields improvements for higher average degrees.

Using this method, for \is{} in graphs with average degree~4, we provide an upper complexity bound
of~$O^*(1.1571^n)$ (Section~\ref{secdeg4}). This bound remains valid for \4is{} outperforming the
best known bound of~$O^*(1.1713^n)$ by~\cite{beigelsoda}.

For \is{} in graphs of average degree~5 we provide a bound
of~$O^*(1.1969^n)$ (Section~\ref{secdeg5}) holding also for \5is{}
and improving the~$O^*(1.2023^n)$ bound
by~\cite{fomin_et_al_soda_06} while, for average degree~6 we obtain
a bound of~$O^*(1.2149^n)$ (Section~\ref{secdeg5}) also improving
the~$O^*(1.2172^n)$ bound by~\cite{fomin_et_al_soda_06}. Note that
for degrees 5 and 6, the results are obtained by a direct
application of the method, without a long case by case branching
analysis.

Let us remark that in the cases of \is{} in graphs of average degree~3 and~4, our bounds
outperform the~$O^*(1.1889^n)$ claimed by~\cite{robson01}.

\section{Graphs of average degree at most 3}\label{secdeg3}

We propose a branch and reduce algorithm for the {\sc maximum
independent set} problem on graphs of average degree at most three.
By local reduction rules and branching, vertices of the input graph
are assigned to be in the computed independent set or not. When a
vertex is decided to be not in the independent set it is removed
from the problem instance, and when a vertex is decided to be in the
independent set it is removed together with all its neighbors.

Given a vertex $v$, we denote $d(v)$ its degree, $N(v)$ its
neighborhood ($v\not\in N(v)$), and $N[v]=N(v)\cup\{v\}$.

\subsection{Simple reduction rules}

Before branching our algorithm applies the following simple reduction rules.
\begin{itemize}
\item If the graph is not connected, recursively solve the problem on each connected component.
            This solves connected components of constant size in constant time.
\item Put isolated vertices in the independent set.
\item Also put any degree 1 vertex in the independent set:
            any independent set containing its neighbor can be modified in one containing the degree 1 vertex of the same size.
\item If for any two adjacent vertices $u$, $v$: $N(u) \subseteq N(v)$, then we say that $u$ \emph{dominates} $v$ and we remove $v$.
            Any maximum independent set containing $v$ can be transformed into another maximum independent set by replacing $v$ by $u$.
\item If there is a vertex $v$ of degree 2 with neighbors $u,w$, we remove $v$ and merge $u$ and $w$.
            This results in a new, possibly higher degree, vertex $x$.
            We refer to this process as \emph{vertex folding}.
            If $x$ is in the computed independent set $I$, then return $(I\setminus\{x\})\cup\{u,w\}$, else return $(I\setminus\{x\})\cup\{v\}$.
            This rule is justified by the fact that if we put any single neighbor of $v$ in $I$ we could equally well have put $v$ itself in $I$.
\end{itemize}
These reduction rules have been thoroughly described in many
publications (\cite{fu3is,3-is} for instance) and therefore need no
further explanation.

\subsection{Small separators}

Following the approach by~\cite{fu3is} we add additional
reduction rules that deal with separators of size~1 and~2. To
prove the worst case time bound we only need these small separators
when one component is of constant size. In this case the recursive
call to the smallest component can be done in constant time.

Let $v$ be an articulation point of $G$ and let $C \subset V$ be the
vertices of the smallest component (vertices in $C$ only have edges
to $v$ or to other vertices in $C$). If the algorithm finds such an
articulation point $v$ it recursively computes a maximum independent
set $I_{\not{v}}$ in the subgraph $G[C]$ and $I_v$ in the subgraph
$G[C\cup\{v\}]$. Notice that $|I_v|$ can be at most~1 larger than
$|I_{\not{v}}|$, and if this is the case then $v \in I_v$. If these
sizes are the same, the algorithm recursively computes the maximum
independent set $I$ in $G[V\setminus(C\cup\{v\})]$ and returns $I
\cup I_{\not{v}}$. This is correct since taking $v$ in the
independent set restricts the possibilities in
$G[V\setminus(C\cup\{v\})]$ more, while it does not increase the
maximum independent set in $C \cup \{v\}$. And if $|I_v| = 1 +
|I_{\not{v}}|$, then the algorithm computes the maximum independent
set $I$ in $G[V \setminus C]$ and returns $I \cup (I_v \setminus
\{v\})$. This is also correct since adding $v$ to $C$ increases the
size of the maximum independent set in $G[C]$ by~1, and this choice
is left to the recursive call on $G[V\setminus C]$.

If the algorithm finds a two separator $\{u,v\}$ of a constant size component $C \subset V$, then it
computes a maximum independent set in the four subgraphs induced by $C$ and any combination of vertices from the separator.
Let $I_{\not{v},\not{u}}$ be the computed maximum independent set in $G[C]$,
$I_{v,\not{u}}$ the computed maximum independent set in $G[C\cup\{v\}]$,
$I_{\not{v},u}$ the computed maximum independent set in $G[C\cup\{u\}]$,
and $I_{v,u}$ the computed maximum independent set in $G[C\cup\{u,v\}]$.
Now consider the following possible cases:
\begin{itemize}
\item $|I_{v,u}|=|I_{\not{v},\not{u}}|+2$, and hence $|I_{v,\not{u}}| = |I_{\not{v},u}| = |I_{\not{v},\not{u}}|+1$.
            The algorithm now computes a maximum independent set in $G[V\setminus C]$ and
            returns $I \cup J$ where $J$ is the set from $\{I_{\not{v},\not{u}},I_{v,\not{u}}, I_{\not{v},u}, I_{v,u}\}$
            which agrees with $I$ on $u$ and $v$.
\item $|I_{v,\not{u}}| = |I_{\not{v},u}| = |I_{v,u}| = |I_{\not{v},\not{u}}|+1$.
            Let $G'$ be $G[V\setminus C]$ with an extra edge added between $u$ and $v$.
            Similar to the previous case, the algorithm computes a maximum independent set in $G'$ and
            returns $I \cup J$, where $J$ is one of the four possible independent sets that agree on $u$ and $v$.
\item $|I_{v,\not{u}}| = |I_{\not{v},\not{u}}|$ and $|I_{\not{v},u}| = |I_{v,u}| = |I_{\not{v},\not{u}}| + 1$ (and the symmetric case).
            $v$ can now safely be discarded since it does not help increasing the size of the independent set in $C\cup\{v\}$.
            The algorithm recursively computes  maximum independent set $I$ in $G[V\setminus (C \cup \{v\}]$
            and returns $I \cup J$, where $J$ is the independent set from $\{I_{\not{v},\not{u}}, I_{\not{v},u}\}$ that agrees on $u$.
\item $|I_{\not{v},u}| = |I_{v,\not{u}}| = |I_{\not{v},\not{u}}|$ and $|I_{v,u}| = |I_{\not{v},\not{u}}| + 1$.
            Let $G'$ be $G[V\setminus C]$ with $u$ and $v$ merged into a single vertex $w$.
            The algorithm makes a recursive call on $G'$ returning $I$.
            If $w \in I$ then we return $9I\setminus\{w\}) \cup I_{v,u}$ and otherwise we return $I \cup I_{\not{v},\not{u}}$.
\item $|I_{v,u}| = |I_{\not{v},u}| = |I_{v,\not{u}}| = |I_{\not{v},\not{u}}|$.
            Now it is safe to use $I_{\not{v},\not{u}}$.
            We make a recursive call on $G[V \setminus (C\cup\{u,v\})]$ resulting
            in $I$ and return $I \cup I_{\not{v},\not{u}}$.
\end{itemize}
In each case we decide whether discarding $u$ and/or $v$ is optimal.
If they cannot be discarded,
we let the recursive call on the larger component decide on their membership of the maximum independent set.

\subsection{Measuring progress}

Let $G=(V,E)$ be a graph with $n$ vertices and $m$ edges. We use
$k=m-n$ as a measure of complexity of the subproblems generated by
our branching algorithm. This means that if our algorithm runs in
$O^*(\gamma^{m-n})$ time, this implies an $O^*(\gamma^{n/2})$
algorithm on connected average degree 3 graphs. Actually, the graph
does not need to be connected; it is just not allowed to have too
much connected components with a negative $m-n$ value. Therefore,
the result applies to any average degree at most three graph that
does not have connected components that are trees. Also notice that
none of the reduction rules (except removing isolated vertices)
increase this complexity measure.

Local configurations of the input graph are considered in order to
decide on the branching. In each branch, a subgraph $G'=(V',E')$ of
$G$ is considered to be removed from the graph after which the
reduction rules are applied again. Let $m'$ be the number of edges
in $G'$, $n'$ be the number of vertices in $G'$, and $e$ be the
number of end points in $G'$ of edges incident to vertices of $G'$
but that are not in $G'$ themselves. In the analysis, we will refer
to these last edges as \emph{external edges}. Note that $G'$ is not
necessarily the subgraph induced by $V'$, {i.e.}, external edges may
be adjacent either to one or to two vertices in $G'$. Removing $G'$
results in a reduction of the complexity measure by at least $m' +
f(e) - n'$. In the ideal case $f(e)=e$ but a few exceptions to this
rule exist, which have to be checked at each branching.  In each
case we look at the number of external edges $e$ that lead to an
overall reduction of the complexity measure. Suppose for example
that we want to take in the solution a vertex $v_1$ which is
adjacent to two degree 3 vertices $v_2$ and $v_3$.
$V'=\{v_1,v_2,v_3\}$ and $E'=\{(v_1v_2),(v_1v_3)\}$ (the edges that
we know for sure). Since $v_2$ and $v_3$ have degree 3, $e=4$. If
$v_2$ is not adjacent to $v_3$ then $f(e)=4$ more edges are removed
when deleting $V'$. On the other hand, if $v_2$ is adjacent to
$v_3$, then only $f(e)=3$ more edges are removed.
\begin{itemize}
\item Some external edges are incident to two vertices in $G'$.
            We refer to this as an \emph{extra adjacency}.
            This can only occur when looking at local configurations larger than a single vertex and its neighborhood (including edges).
            Adding these edges to $G'$ results in $e$ being reduced by 2, while the complexity measure only decreases by 1.
\item After removing $G'$ from $G$, a number of connected components arise some of which are trees.
            A tree has complexity $-1$ and is completely removed by the reduction rules.
            Let $t$ be the number of external edges incident to such a tree ($t\geq 3$ since reduction rules produce a graph
            of minimum degree at least 3).
            For each tree that we add to $G'$ we decrease $e$ by $t$ while increasing the complexity measure by only $t-1$.
            So in the worst case $e$ is decreased by 3 and the complexity measure is decreased by 2.
\item A special case arises when $G'$ is the neighborhood of a vertex $v$ and there are no 4-cycles in the graph.
            In this case there can be no induced trees because after the removal of $G'$ all vertices are of degree at least two
            and hence $f(e)=e$.
\end{itemize}

\subsection{Induced trees}\label{sectreedeg3}

In order to prevent tree components from being created,
we add some additional reduction rules and discuss some cases in which no trees can arise.

When discarding a single vertex, no tree can be created since vertex
folding causes all vertices in the instance graph to be of degree at
least three. When taking a single vertex $v$ in the independent set
and discarding all its neighbors, several cases can arise. If $v$ is
of degree more than three, these cases are handled with the
description of the branching. In this section, we treat the cases
where $v$ is of degree 3 in a maximum degree 4 graph and distinguish
on the number of vertices in an induced tree.

Let $a,b,c$ be the neighbors of $v$ and notice that they all have at
least one edge not incident to $v$ or the tree $T$ (otherwise there
exists a small separator). If the tree $T$ consists of a number of
vertices equal to:
\begin{enumerate}
\item $T$ is a single degree 3 vertex. We consider the following possibilities:
    \begin{itemize}
    \item There is an edge in $N(v)$. It is now optimal to take $v$ and $T$ in the independent set.
                We can take only two from $a$, $b$ or $c$ from which any one causes $v$ and $T$ to be discarded,
                while taking $v$ and $T$ poses less restrictions on the remaining graph.
    \item 
                Notice that either we take only two vertices among $a,b,c,v,t$ and in this case taking $v$ and $T$ is optimal,
                or we take three vertices and the only possibility is to take $a$, $b$ and $c$.
                We postpone this choice, but reduce the instance by removing $v$ and $T$ and merging $a$, $b$ and $c$ to a single vertex.
    \end{itemize}
\item $T$ consists of two adjacent degree 3 vertices.
    Now at least one vertex is adjacent to both tree vertices.
    Let this be $a$; now $b$ and $c$ are adjacent to one or both tree vertices.
    \begin{itemize}
    \item $b$ and/or $c$ is adjacent to both tree vertices.
                In this case, one of the tree vertices dominates the other and this reduction rule fires.
    \item $b$ and $c$ are of degree 3.
                This generalizes the 1-tree case: after taking $a$ it is optimal to take $b$ and $c$, and
                after discarding $a$ it is optimal to take $v$ and any one vertex of $T$.
                Hence, we again remove $v$ and $T$ and merge $a$, $b$ and $c$ to a single vertex.
    \end{itemize}
\item $T$ consists of three vertices.
    At least two neighbors of $v$, say $a$ and $b$, have two tree neighbors (Figure~\ref{figA43}).
    Consider the the maximum independent set $I'$ in $G[N[v]\cup T]$.
    If $a \in I'$ and hence it's neighbors are not, only $b,c$ and one vertex from $T$ remain from which by adjacencies to $T$
    only two can be in $I'$.
    The same goes with $a$ and $b$ switched.
    And if we discard $a$ and $b$, it is clear that it is optimal to pick $v$ and two vertices from $T$ while discarding $c$.
    Over all three cases, the last never gives a smaller independent set in $G[N[v]\cup T]$, while posing the fewest restrictions
    on the rest of the graph; therefore we let our algorithm pick these vertices.

\begin{figure}[thb]
\begin{center}
\includegraphics{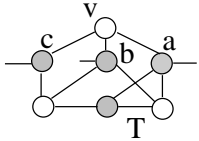}
\caption{$T$ consists of three vertices and at least two neighbors of $v$,~$a$ and~$b$, have two tree neighbors.}\label{figA43}
\end{center}
\end{figure}

\item $T$ consists of four vertices.
    Now all neighbors of $v$ are of degree 4.
    By a similar argument, it is optimal to pick $v$ and a maximum independent set from $G[T]$.
    Depending on the shape of $G[T]$, this independent set can be of size two or three.
\end{enumerate}
In a maximum degree 4 graph, the only remaining cases are when $v$
has multiple non-adjacent degree 4 neighbors and these trees consist
of one or two vertices. In these cases, the creation of trees is
handled with the description of the branching.

We will later refer to a created tree components consisting of $k$ vertices as a $k$-tree.

\subsection{Branching on non-3-regular graphs}

The worst case of our algorithm arises when the graph $G$ is 3-regular.
In this section, we describe the branching of our algorithm when this is not the case.
Observe that vertex folding can produce non-3-regular graphs after deciding for a vertex $v$ in
a 3-regular graph whether $v$ goes in the maximum independent set or not.
This observation is used later.
Therefore, we need the following lemma.

\begin{lem}\label{lemdeg3}
Let $T(k)$ be the number of subproblems generated when branching on a graph $G$ of complexity $k$.
If $G$ is not 3-regular then either:
\begin{enumerate}
\item $G$ has a vertex of degree at least five and $T(k) \leq T(k-4) + T(k-7)$.
\item $G$ has a vertex of degree 4 that is part of a triangle or 4-cycle also containing at least one degree 3 vertex,
            and there are no triangles or 4-cycles containing only degree 3 vertices,
            then: $T(k) \leq T(k-5) + T(k-6)$ or $T(k) \leq 2T(k-8)+2T(k-12)$.
\item $G$ has a vertex of degree 4 that is part of a triangle containing at least one degree 3 vertex,
            and there is no constraint on the degree 3 vertices, then: $T(k) \leq T(k-4) + T(k-6)$ or $T(k) \leq 2T(k-8)+2T(k-12)$.
\item $G$ has at least one vertex of degree 4, none of which satisfy the previous case, and $T(k) \leq T(k-3) + T(k-7)$.
\end{enumerate}
Or a better branching exists.
\end{lem}
When referring to this lemma, often only the branching behavior of
its worst case (case 4) is used in the argument.

Before proving the lemma in a step by step fashion, we need the
concept of a \emph{mirror} (\cite{fomin_et_al_soda_06}). A vertex $m
\in V$ is a mirror of $v \in V$ if $N(v) \setminus N(m)$ forms a
clique. Mirrors are exploited by our algorithm in the following way:
whenever we branch on $v$ and discard it at least two of the
neighbors of $v$ should be in the maximum independent set. Namely,
if we take only one, we could equally well have picked $v$ which is
done in the other branch. Since we can take only one vertex form the
clique $N(v) \setminus N(m)$, a vertex from $N(v) \cap N(m)$ must be
in the independent set. Hence we can safely discard $m$ also without
changing the size of the maximum independent set.

Notice that the only 4-cycles in a maximum degree 4 graph in which
no degree 3 vertex has a mirror consists of four degree 4 vertices.
These facts are exploited when we try to limit the number of tree
components created by branching.

For the proof we also need the general observation that for any $T(k-r_1) + T(k-r_2)$ branch with $r_1 < r_2$,
a $T(k-r_1-c) + T(k-r_2+c)$ branch is a better branch as long as $r_1+c \leq r_2-c$.

The proof will be divided over several subsections corresponding to the various local configurations to which the lemma applies.

\subsubsection{Vertices of degree at least five}\label{subsecvertex5}

\begin{figure}[thb]
\begin{center}
\includegraphics{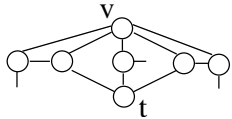}
\caption{Vertices of degree at least five.}\label{figA51}
\end{center}
\end{figure}

Let $v$ be a vertex of degree at least
five (Figure~\ref{figA51}). Our algorithm branches by either taking $v$ in the independent
set and discarding $N(v)$ or discarding $v$. If $v$ is discarded,
one vertex is removed and at least five edges are removed forming a
subproblem of complexity no more than $k-4$. If $v$ is put in the
independent set, $N[v]$ is removed. In the worst case all neighbors
of $v$ have degree 3. By domination all vertices in $N(v)$ have at
least one neighbor outside of $N[v]$. Together this leads to at most
two edges in $G[N(v)]$ and at least six external edges. If no trees
are created these 6 edges and the 7 edges in $G[N[v]]$ minus 6
vertices lead to the required size reduction of $k-7$. And if any
neighbor of $v$ has degree 4 or more, or there are fewer edges in
$G[N(v)]$, then the number of external edges is large enough to
guarantee this size reduction of $k-7$.

What remains is to handle the special case where all vertices in
$N(v)$ are of degree 3, there are six external edges, and a tree is
created. This tree will be a single degree 3 vertex $t$, since
otherwise there exists a two separator in $N(v)$. Notice that $v$ is
a mirror of $t$. We branch on $t$. Taking $t$ leads to the removal
of 4 vertices and 9 edges: $T(k-5)$. And discarding $t$ and $v$
leads to the removal of 8 edges and two vertices: $T(k-6)$. In the
last case there again can be trees, but this implies that the entire
component is of constant size. This branching with $T(k) \leq T(k-5)
+ T(k-6)$ is better than the required $T(k) \leq T(k-4) + T(k-7)$.

\subsubsection{Triangles with two degree 4 vertices and a degree 3 vertex}

Let $x$, $y$, $w$ be a triangle (3-cycle) in the graph with $d(x)=d(y)=4$ and $d(w)=3$, also let $v$ be the third neighbor of $w$.
Notice that discarding $v$ causes domination which results in $w$ being taken in the maximum independent set.
Our goal is to show that there always exist an efficient enough branching.

If $v$ is of degree 4, discarding $v$ and taking $w$ leads to the
removal 11 edges and 4 vertices: $T(k-7)$. Notice that tree
components cannot be created because these would have been removed
by the preprocessing since there is an edge in $G[N(w)]$. Taking $v$
and removing $N[v]$ results in the removal of 3 edges incident to
$w$ and at least 8 more edges and 5 vertices. If in this last case
all neighbors of $v$ are of degree 3, then there are at most 6
external edges and hence there can be at most one tree. Otherwise
any degree 4 neighbors of $v$ cause even more edges to be removed,
compensating for any possible tree. This results in $T(k-5)$: $k-6$
with a $+1$ for the tree.

\begin{figure}[thb]
\begin{center}
\includegraphics{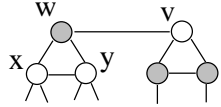}
\caption{Vertex~$v$ has degree~3.}\label{fig452}
\end{center}
\end{figure}

If $v$ is of degree 3 (Figure~\ref{fig452}), discarding $v$ and taking $w$ leads to the
removal of of at least 10 edges and 4 vertices: $T(k-6)$. Now if
also $v$ is not part of any triangle or has a degree 4 neighbor
(case 2 of the lemma) taking $v$ removes 9 edges and 4 vertices:
$T(k-5)$. And if $v$ is part of a triangle of degree 3 vertices
(case 3 of the lemma) taking $v$ removes 8 edges and 4 vertices
$T(k-4)$.

\subsubsection{Triangles with one degree 4 vertex and two degree 3 vertices}

\begin{figure}[thb]
\begin{center}
\includegraphics{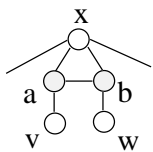}
\caption{Triangles with one degree 4 vertex and two degree 3 vertices.}\label{fig453gen}
\end{center}
\end{figure}

When there is only one degree 4 vertex, the situation gets a lot
more complicated. Let $x$, $a$ and $b$ be the triangle vertices with
$d(x)=4$ and $d(a)=d(b)=3$, also let $v$ be the third neighbor of
$a$, and let $w$ be the third neighbor of $b$ (Figure~\ref{fig453gen}). $v$ and $w$ are not
adjacent to $x$ and $v\not=w$ by domination. If $v$ and $w$ are
adjacent, we can safely discard $x$ reducing the graph. This last
fact follows from the fact that if we pick $v$ we would also pick
$b$, and if we discard $v$, its mirror $b$ is also discarded which
results in $a$ being picked. In both cases a neighbor of $x$ is in a
maximum independent set and hence $x$ can safely be discarded. So we
assume that $v$ and $w$ are non-adjacent.

If $v$ or $w$, say $v$, is of degree 4, taking $v$ removes at least
11 edges and 5 vertices, but since there are 6 external edges there
can be a tree: $T(k-5)$. And if there are more external edges (less
edges in $N(v)$) the number of edges removed increases. Discarding
$v$ and by domination taking $a$ leads to the removal of 10 edges
and 4 vertices: $T(k-6)$. Although in the last case $a$ is a degree
3 vertex with two degree 4 neighbors, there cannot be any trees
since there is an edge in $G[N(a)]$: a tree would fire a reduction
rule for trees. So from now on we can assume that $v$ and $w$ are of
degree 3.

\begin{figure}[thb]
\begin{center}
\includegraphics{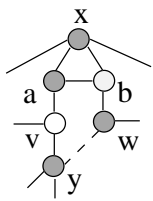}
\caption{Vertex~$v$, has a degree 4 neighbor~$y$.}\label{fig453b}
\end{center}
\end{figure}

Consider the case where $v$ or $w$, say $v$, has a degree 4 neighbor
$y$ (Figure~\ref{fig453b}). Suppose that $y$ does not form a triangle with $v$, then taking
$v$ removes at least 10 edges and 4 vertices: $T(k-6)$. Discarding
$v$ and by domination taking $a$ removes at least 9 edges and 4
vertices: $T(k-5)$. If $y$ does from a triangle with $v$ we branch
on $w$. If $w$ has a degree 4 neighbor or is not involved in a
triangle (case 2 of the lemma), then taking $w$ results as before in
$T(k-5)$. Discarding $w$ by domination results in taking $b$ which
again by dominating results in taking $v$. In total 15 edges are
removed from which 7 external edges and 7 vertices. Because of the
separators there can be at most 2 extra adjacencies in the worst
case leaving 3 external edges and $T(k-6)$. Note that trees are
beneficial over extra adjacencies. This leaves the case where $w$
has only degree 3 neighbors with which it forms a triangle (case 3
of the lemma). In this case taking $w$ only leads to $T(k-4)$, and
$T(k) \leq T(k-4) + T(k-6)$ is enough. So we can assume $v$ and $w$
to be of degree 3 and have no degree 4 neighbors.

\begin{figure}[thb]
\begin{center}
\includegraphics{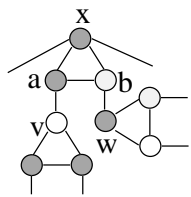}
\caption{Vertex~$v$ is part of a triangle.}\label{fig453c}
\end{center}
\end{figure}

Suppose that $v$ or $w$, say $v$, is part of a triangle (Figure~\ref{fig453c}).
Notice that we are now in case 3 of the lemma.
We branch on $w$.
If we take $w$ the worst case arises when $w$ is also part of a triangle; 8 edges and 4 vertices are removed: $T(k-4)$.
And if we discard $w$ by domination $b$ and $v$ are put in the independent set removing a total of at least 14 edges
from which 6 external and 7 vertices.
Because of the small separator rules, the external edges can form at most one extra adjacency or tree leading to $T(k-6)$.
So at this point we can also assume that $v$ and $w$ are not part of any triangle.

\begin{figure}[thb]
\begin{center}
\includegraphics{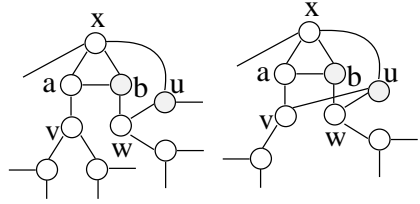}
\caption{Vertex~$w$, has a neighbor $u\not=a,b$ that is adjacent to~$x$.}\label{fig453d}
\end{center}
\end{figure}

Suppose $v$ or $w$, say $w$, has a neighbor $u\not=a,b$ that is
adjacent to $x$ (Figure~\ref{fig453d}). We branch on $v$ and if we discard $v$, $a$ is
picked by domination and we still have $T(k-5)$. If we take $v$ we
have the situation that $b$ becomes a degree 2 vertex which
neighbors $x$ and $w$ are folded to a single vertex. Notice that
both $x$ and $w$ are adjacent to $u$ and hence this folding removes
an additional edge: $T(k-6)$. The only case in which the above does
not holds is when $v$ and $w$ are both a neighbor of $u$. We reduce
this exceptional case by noting that a tree reduction rule fires
when considering branching on $u$ (without actually branching on $u$
of course). This is the rule dealing with $u$ having one degree 4
neighbor and a 2-tree $\{a,b\}$. Hence, now we can also assume that
$v$ and $w$ have no neighbors besides $a$ and $b$ that are adjacent
$x$.

We conclude this subsection by describing three more branches depending on the number of vertices in $X = (N(w) \cup N(v)) \setminus \{a,b\}$.

\begin{figure}[thb]
\begin{center}
\includegraphics{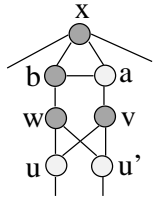}
\caption{Vertices~$v$ and~$w$ are adjacent to both of~$u$ and~$u'$.}\label{fig453e2}
\end{center}
\end{figure}

Assume that there exist two vertices~$u$ and~$u'$ such that~$v$ and~$w$ are adjacent to
both of them (Figure~\ref{fig453e2}). Notice that if we take $v$ in the independent set it is
optimal to also pick $w$ and vice versa. Hence we branch, taking
both $v$ and $w$ or discarding both. If we take both $v$ and $w$, 11
edges are removed and 6 vertices: $T(k-5)$. If we discard both $v$
and $w$ we can take $a$ in the independent set and remove 11 edges
and 5 vertices: $T(k-6)$. When taking both $v$ and $w$ there can be
not trees since there are only 4 external edges. When discarding
both $v$ and $w$ two tree leaves $u$ and $u'$ are formed, but they
cannot form a tree since their adjacency results in a one separator,
and adjacency to the only possibly degree 2 vertices (neighbors of
$x$) results in a constant size component or a small separator. Also
there cannot be any extra adjacencies because then there exists a
small separator.

\begin{figure}[thb]
\begin{center}
\includegraphics{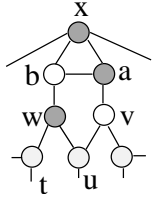}
\caption{The case $|X|=3$.}\label{fig453e3}
\end{center}
\end{figure}

If $|X|=3$, let $u \in X$ be the common neighbor of $v$ and $w$ and let $t \in X$ be the third neighbor of $w$ (Figure~\ref{fig453e3}). We branch on $t$. If we take $t$ in the independent set we also take $b$ by domination. This
results in the removal of 7 vertices and 15 edges if $t$ has a
degree 4 neighbor or there is no triangle involving $t$, otherwise
only 14 edges are removed. Since there can be at most 8 external
edges with this number of removed edges, and hence at most 2 extra
adjacencies or trees we have $T(k-6)$ or $T(k-5)$. If we discard
$t$, 3 edges and 1 vertex are removed and the folding of $w$ results
in a new degree 4 vertex $[bu]$. This new vertex can be discarded
directly since it is dominated by $a$ resulting in an additional
removal of 4 edges and 1 vertex. This leads to $T(k-5)$ in total.
Furthermore, there cannot be any induced trees since there can be at
most one vertex of degree less than two (adjacent to $t$ and $u$,
but no to $w$) which cannot become an isolated vertex. Depending on
whether $t$ is in a triangle we are in case 2 or 3 or the lemma and
we have a good enough branching.

If $|X|=4$, all neighbors of $v$ and $w$ are disjoint. We branch on
$v$. If we take $v$, we remove 9 edges and 4 vertices, and if we
discard $v$, we take $a$ and again remove 9 edges and 4 vertices.
Also notice that if we take $v$, $b$ is folded resulting in a degree
four vertex $[xw]$. And if we take $a$, $w$ is folded resulting in
the removal of an extra edge if its neighbors have another common
neighbor or also in a degree 4 vertex. In the first case we have
$T(k) \leq T(k-5) + T(k-6)$, and in the second case we inductively
apply our lemma to both generated branches. This leads to the $T(k)
\leq 2T(k-8) + 2T(k-12)$ in the lemma.

Remark that $T(k) \leq T(k-5) + T(k-5)$ has a smaller solution than
$T(k) \leq 2T(k-8) + 2T(k-12)$. However, after a bad branch in a
3-regular graph the second gives a better solution when applied to
one of both branches. This is because it is a composition of three
branchings that are all a lot better than the bad 3-regular graph
branching.

\subsubsection{4-cycles in which a degree 4 vertex is a mirror of a degree 3 vertex}

\begin{figure}[thb]
\begin{center}
\includegraphics{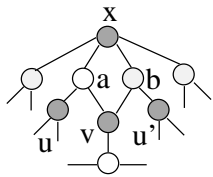}
\caption{Vertex~$x$ is a degree~4 vertex that is a mirror of the degree three vertex~$v$.}\label{fig454}
\end{center}
\end{figure}

Let $x$ be the degree 4 vertex that is a mirror of the degree three
vertex $v$, let $a$ and $b$ be their common neighbors, and let $w$
be the third neighbor of $v$ (Figure~\ref{fig454}). If we branch on $v$ and take $v$, we
remove at least 9 edges and 4 vertices, and when we discard $v$ and
also $x$ because it is a mirror of $v$, we remove 7 edges and 2
vertices: $T(k) \leq T(k-5) + T(k-5)$. We show that in any case we
can always find an extra complexity reduction in one of both
branches leading to the required result. Notice that if we discard
$v$ and $x$, there can be no trees since the only possible leaves
created are $a$ and $b$. These two vertices may not be adjacent by
dominance. And if they form a tree with any vertex that used to be
adjacent to $x$ or $v$, there would have existed a small separator
or there is no tree at all. Also, any extra adjacency results in
triangles involving degree 3 and four vertices which are handled in
the previous subsection.

First assume that $a$, $b$ or $w$ is of degree 4, then $T(k-6)$ when
taking $v$. Since $v$ is of degree 3 there can only be trees if two
or more vertices from $\{a,b,w\}$ are of degree 4. But in this case
even more edges are removed, because $a$, $b$ and $w$ are
non-adjacent to each other, which compensates for the creation of a
tree. So we can assume that $a$, $b$ and $w$ are of degree 3.

If both $a$ and $b$ have another common neighbor $y \not= v$, then
the graph can be reduced without branching. Indeed, among
$a,b,v,x,y$, in an optimum solution either we take 2 vertices (hence
$a,b$) or three vertices (hence $v,x,y$). We can replace the
subgraph induced by $a,b,v,x,y$ by one vertex that we link to the
other neighbors of $v,x,y$. So we can assume that $a$ and $b$ do not
have more than two common neighbors.

Let $u$ and $u'$ be the third neighbors of $a$ and $b$,
respectively. When discarding $v$ and $x$, both $a$ and $b$ are
taken in the independent set and $u$ and $u'$ are discarded also.
This means that 13 edges form which 7 external edges and 6 vertices
are removed. First assume that $u$ and $u'$ are vertices of degree
three. The only possible adjacencies are those between $u$ and $u'$,
or $u$ or $u'$ and $v$. But there can be only one adjacency because
if we take two we have a small separator. So we end up removing 12
edges from which 5 external edges and 6 vertices which cannot create
trees: $T(k-6)$. Now suppose that $u$ or $u'$ is of degree 4 and
notice that the extra edges removed compensate for any possible
extra adjacencies or created tree components.

\subsubsection{4-cycles that contain degree 3 and 4 vertices, while no degree 4 vertex is a mirror of a degree 3 vertex}

This can only be the case if the cycle consists of two degree 4
vertices $x$, $y$ and two degree 3 vertices $u$, $v$ with $x$ and
$y$ not adjacent. There are no other adjacencies than the cycle
between these vertices by cases presented in previous subsections.

\begin{figure}[thb]
\begin{center}
\includegraphics{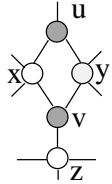}
\caption{Vertex~$v$, has a third degree~4 neighbor~$z$.}\label{fig455}
\end{center}
\end{figure}

Suppose that either $u$ or $v$, say $v$, has a third degree 4
neighbor $z$ (Figure~\ref{fig455}). Notice that this neighbor cannot be adjacent to $x$ or
$y$. If we branch on $v$ and take $v$, we remove 12 edges and 4
vertices, and if we discard $v$ and its mirror $u$ we remove 6 edges
and 2 vertices. So if no trees are created, we have: $T(k) \leq
T(k-8) + T(k-4)$. Because of the reduction rules a tree can consist
of at most two vertices. Also at most one tree can be formed,
otherwise it would be optimal to pick $v$ and a maximum independent
sets in each created tree. Consider both cases:
\begin{enumerate}
\item A tree consisting of 1 vertex. Since this one vertex is a degree 3 vertex and a mirror we assume without loss of generality that it is $u$.
            Taking $v$ now leads to $T(k-7)$, while discarding it results in $T(k-4)$ and $x$, $y$ and $z$ to be of degree 2.
            If any of these vertices are in another 4-cycle after discarding $u$ and $v$, folding causes an additional edge to be removed: $T(k)\leq T(k-5)+T(k-7)$.
            And if not, degree 4 vertices are created because $x$, $y$ and $w$ are non-adjacent.
            We inductively apply the lemma to this case and obtain $T(k) \leq 2T(k-7)+T(k-11)$.
\item Trees consisting of 2 vertices.
            If any vertex in the tree is of degree 4 it dominates the other.
            So both vertices are of degree 3.
            But then a vertex from $x$, $y$ and $w$ forms a triangle with this tree which was covered by branching rules in previous subsections.
\end{enumerate}
All the branching rules described above give a better bound on the
running time of our algorithm than required by the lemma. So we can
assume that $u$ and $v$ have no degree 4 neighbors not on the
4-cycle.

Let $w$ be the degree 3 neighbor of $v$. It is not adjacent to $u$
since that would imply that $x$ and $y$ are mirrors of $w$. Since
the $w$ cannot be adjacent to $x$ or $y$, taking $v$ results in the
removal of 11 edges and 4 vertices: $T(k-7)$. Discarding $v$ and $u$
leads to the removal of 6 edges and 2 vertices: $T(k-4)$. Because in
the last case $x$ and $y$ will be folded and they are not adjacent
to other created degree 2 vertices (then there would be triangles
involving degree 3 and four vertices), a vertex of degree at least
four is created or at least one additional edge is removed. This
again leads to $T(k)\leq T(k-5)+T(k-7)$ or $T(k) \leq
2T(k-7)+T(k-11)$ by applying the lemma inductively. We do require
here that there are no trees created. But If a tree is created we
follow the above reasoning: this can only be a single tree
consisting of one vertex (two vertices lead to triangles with degree
and four vertices) and there can be only one such tree. In this case
$w$ is adjacent to the tree and to $v$ and therefore has $x$ and $y$
as a mirror and we refer to the previous subsection.

\subsubsection{A degree 4 vertex that is not involved in any triangle or 4-cycle with any degree 3 vertex}
Let $x$ be this vertex. If all its neighbors are of degree 3,
branching on it results in $T(k) \leq T(k-7)+T(k-3)$. In this case
there cannot be any created trees for any tree leaf is of degree at
least three before branching and therefore must have at least two
neighbors in $N(x)$ to become a leaf. But in this last case, there
exist four cycles with degree 3 and four vertices on it which
contradicts our assumption.

If $x$ has degree 4 neighbors, the number of edges removed increases
and there can still be no trees unless at least three neighbors of
$x$ are of degree 4 and every tree leaf vertex originally was a
degree 4 vertex. If $x$ has three neighbors of degree 4 there are at
least 13 edges removed, in which case there are 7 external edges.
This can lead to at most one tree and $T(k-7)$ as required. If there
are more external edges, there will also be more edges removed
keeping this reduction. Finally if $x$ has four degree 4 neighbors,
we remove at least 12 edges from which 4 external edges again
leading to $T(k-7)$. Here any tree implies more external edges and
hence more edges removed also keeping this reduction.

Putting all the above together completes the proof of Lemma~\ref{lemdeg3}.

\subsection{Branching on 3-regular graphs with triangles or 4-cycles}

Whenever the algorithm encounters a 3-regular graph that contains triangles or 4-cycles we can still do better than our worst case.
This is settled by a second lemma.
\begin{lem}\label{lem2}
Let $T(k)$ be the number of subproblems generated when branching on a graph $G$ of complexity $k$.
If $G$ is 3-regular and contains a triangle or 4-cycle, then $T(k) \leq T(k-4)+T(k-5)$ or a better branching exists.
\end{lem}
We will now prove this lemma.

\subsubsection{3-regular graphs that contain a triangle}

Let $a$, $b$, $c$ be the triangle vertices.
Assume that one of these three vertices, say $a$, has a neighbor $v$ not in any triangle in the graph.
The algorithm branches on $v$.
If $v$ is included in the independent set, 9 edges and 4 vertices are removed: $T(k-5)$.
And if $v$ is discarded and by domination $a$ is put in the independent set, 8 edges and 4 vertices are removed: $T(k-4)$.

This gives the required branching unless all three triangle vertices
only have neighbors that also form triangles. In that case we branch
on $a$. If $a$ is discarded, domination forces $v$ in the
independent set which symmetric to the above resulting in $T(k-4)$.
When $a$ is included in the independent set, $b$ and $c$ are
discarded which by domination results in the third neighbors of $b$
and $c$ to be put in the independent set. Now a total of 18 edges
from which 6 external edges and 10 vertices are removed. Adding the
at most one extra adjacency or tree this results in $T(k-7)$ which
is more than enough.

\subsubsection{Triangle free 3-regular graphs that contain a 4-cycle}
Let $v$ be a vertex on the 4-cycle. Observe that vertices opposite
to $v$ on a 4-cycle are mirrors of $v$. If we branch on $v$,
triangle freeness results in the removal of 9 edges and 4 vertices
when taking $v$: $T(k-5)$. When discarding $v$, its mirrors can also
be discarded resulting in the removal of 6 edges and 2 vertices if
$v$ has only one mirror and possibly more if $v$ has two or three
mirrors: $T(k-4)$. Notice that two degree 1 vertices are formed that
are not part of a tree. This is because their adjacency implies
domination, and if they are adjacent to degree 2 vertices a small
separator exists. When $v$ has more than one mirror, single vertex
trees can be created in $N(v)$. These extra mirrors compensate more
than enough to maintain our $T(k-4)$.

The proof of Lemma~\ref{lem2} is now completed.

\subsection{Branching on 3-regular graphs without triangles or 4-cycles}

Having gone through enough preparation, we are now ready for the third lemma on the branching behavior of our algorithm.
Taken together, these lemmata will directly result in the claimed running time.
\begin{lem}\label{lem3}
Let $T(k)$ be the number of subproblems generated when branching on a graph $G$ of complexity $k$.
If $G$ is 3-regular and contains no triangles or 4-cycles, then branching on any vertex results in $T(k) \leq T_2(k-2) + T_4(k-5)$,
where $T_2$ and $T_4$ correspond to situations 2 and 4 from lemma 1, respectively, or a better branching exists.
\end{lem}
This leads to the worst case recurrence relation $T(k) \leq
T(k-8)+2T(k-10)+T(k-12)+2T(k-14)$ and a running time of
$O^*(1.17802^k)$.

Taking $v$ in the independent set results in $T(k-5)$, and discarding $v$ results in $T(k-2)$.
Clearly this branching is not good enough and we will show that we can always do better.

Before we consider the subcases involved in this lemma, observe what
happens when branching on $v$. Let $x$, $y$, $z$ be the neighbors of
$v$. Because of triangle and 4-cycle freeness they have disjoint
neighbors; let $N(x) = \{v,a,b\}$, $N(y) = \{v,c,d\}$ and $N(z) =
\{v,e,f\}$. Notice that there cannot be any adjacencies within these
neighborhoods, but there can be adjacencies between $a, \ldots, f$
if they are neighbors of different vertices in $N(v)$. When $v$ is
discarded, these neighborhoods ($N(x)$, $N(y)$ and $N(z)$) are
merged to single vertices. Their degrees and relative positions in
the reduced graph depends on the adjacencies between vertices in
these neighborhoods. Consider the different possible number of
adjacencies; we number cases to deal with later:
\begin{enumerate}
\setcounter{enumi}{-1}
\item If there is no adjacency between $N(x)$, $N(y)$ and $N(z)$, each neighborhood is merged to a degree 4 vertex
            none of which are adjacent in the reduced graph when discarding $v$ (1).
\item If there is one adjacency between $N(x)$, $N(y)$ and $N(z)$, discarding $v$ results in three degree 4 vertices
            only two of which are adjacent (2).
\item If there are two adjacencies between $N(x)$, $N(y)$ and $N(z)$,
            these can either be between the same neighborhoods or involving all three neighborhoods.
            In the first case, an extra edge is removed because the merged vertices cannot have two edges between them.
            This results in their degrees to be only three, while the other neighborhood is merged to a non-adjacent degree 4 vertex (4).
            In the second case, we have three degree 4 vertices from which one is adjacent to the other two but they do not forming a triangle.
            We will call this a path of three degree 4 vertices (3).
\item If there are three adjacencies between $N(x)$, $N(y)$ and $N(z)$,
            either there are multiple adjacencies between the neighborhoods as in the previous case resulting in the removal of an extra edge (5),
            or a clique of three degree 4 vertices is formed (6).
\item If there are four adjacencies between $N(x)$, $N(y)$ and $N(z)$,
            there are either two double adjacencies resulting in two additional edges being removed and
            $T(k) \leq T(k-4) + T(k-5)$, or a single double adjacency and two single adjacencies.
            In the second case these adjacencies result in two folded degree 3 vertices forming a triangle with a degree 4 vertex.
            Here we can apply case 3 of Lemma~\ref{lemdeg3} obtaining: $T(k) \leq T(k-5) + T(k-3-4) + T(k-3-6) = T(k-5) +T(k-7)+T(k-9)$.
\item If there are five adjacencies between $N(x)$, $N(y)$ and $N(z)$, we have a two separator and are done.
\item If there are six adjacencies between $N(x)$, $N(y)$ and $N(z)$, we have a constant size component and are done too.
\end{enumerate}
Notice that these adjacencies also have meaning when taking $v$ in
the independent set. Namely, if these neighborhoods are
non-adjacent, triangle and 4-cycle freeness also ensures the
creation of degree 4 vertices after taking $v$. However, if for
example $a$ and $f$ are adjacent, then taking $v$ results in these
vertices to become two adjacent degree 2 vertices. In this case,
these vertices are merged resulting in nothing more than an edge
between their other neighbors replacing the old edges from these
neighbors to $a$ and $f$. In the case of three adjacencies without
double adjacencies (6), this can very well lead to a new 3-regular
graph without triangles or 4-cycles. In any other case, we can apply Lemma~\ref{lemdeg3} also to the branch in which we take $v$ since a degree 4
vertex is formed This is the $T_4(k-5)$ term in the lemma.

The six numbered cases are handled in more detail in the rest of
this section. We know that in each case the reduced graph after
discarding $v$ has at most three degree 4 vertices; all other
vertices are of degree 3. Because the graph is triangle and 4-cycle
free before applying this lemma, a new triangle or 4-cycle created
after discarding $v$ must involve the vertices obtained by folding.
And, if any of the degree 4 vertices form a triangle or 4-cycle with
any degree 3 vertex, we apply Lemma~\ref{lemdeg3}. If no degree 3 vertices are
created by folding, this results in the required branch of
$T_2(k-2)$, otherwise at least one extra edge is removed and we need
case 3 of Lemma~\ref{lemdeg3} resulting in even better branches:
$T(k-3-4)+T(k-3-6)$. Therefore, we can assume that no triangles nor
4-cycles involving both degree 3 and four vertices exist.

\subsubsection{Three non-adjacent degree 4 vertices}

Following the reasoning for the general case, we apply Lemma~\ref{lemdeg3} to the case where we take $v$.
A $T(k) \leq T_4(k-3) + T(k-9)$ branch applied to the graph of complexity $k-2$ after discarding $v$,
where $T_4(k-3)$ means we apply Lemma's~1 case 4 also here,
leads to $T(k) \leq 2T(k-8) + T(k-11) + 2T(k-12)$ which is sufficient.

\begin{figure}[thb]
\begin{center}
\includegraphics{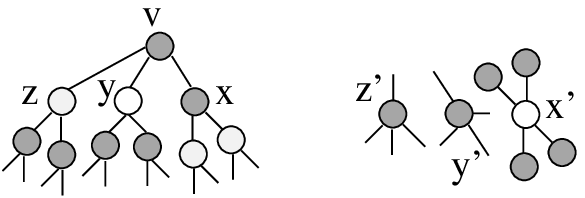}
\caption{}\label{fig71}
\end{center}
\end{figure}

The $T(k) \leq T_4(k-3) + T(k-9)$ branch follows from exploiting a
little bit more information we have about the maximum independent
set we need to compute in this branch than just the reduced graph.
This reasoning is quite similar to exploiting mirrors. Namely, if
$v$ is discarded we know that we need to pick at least two of the
three neighbors of $v$: if we pick only one we could equally well
have taken $v$ which is done in the other branch already. This
observation becomes slightly more complicated because we just folded
the neighbors of $v$. Consider the vertex $x'$ that is the result of
folding vertex $x$. The original vertex $x$ is taken in the
independent set if and only if $x'$ is discarded in the reduced
graph. So, the fact that we needed to pick at least two vertices
from $N(v)$ results in us being allowed to pick at most one vertex
from the three degree 4 vertices created by folding the neighbors of
$v$. Hence, picking any vertex from the three folded vertices allows
us to discard the other two. The above discussion is illustrated in Figure~\ref{fig71}.

Let $x'$, $y'$ and $z'$ be the degree 4 vertices resulting from
folding $x$, $y$ and $z$, respectively. If we discard $x'$, we
remove 4 edges and 1 vertex. Moreover, after discarding $x'$, at
least one degree 4 vertex remains in the graph resulting in
$T_4(k-3)$, or at least one extra edge is removed by folding
resulting in $T(k-4)$ which in this case is even better.

When we take $x'$, which has four degree 3 neighbors, we can discard
these and both $y'$ and $z'$ resulting in the removal of 20 edges
from which 16 external edges and 7 vertices. Because $y'$ and $z'$
are non adjacent and they can only be adjacent to a single neighbor
of $x'$ (or a 4-cycle would exist), there are at most two extra
adjacencies. In this case there are 12 external edges left, but
these can only form very specific trees leading to
$T(k-20+7+2+2)=T(k-9)$. This is because every tree vertex $t$ can
only have neighbors that are distance 3 away from each other in
$G[V\setminus\{t\}]$ because of the triangle and 4-cycle freeness.
The only 1-trees that can be created are adjacent to both $y'$ and
$z'$ and a neighbor of $x'$ that is not adjacent to either $y'$ or
$z'$. There can be at most one such trees, since two 1-trees
adjacent to two of the same vertices also create a 4-cycle. And, it
can only exist if $y'$ and $z'$ are adjacent to different neighbors
of $x'$. This results in 9 remaining external edges that because of
the small separators can form only one larger tree. If there is no
1-tree, larger trees use more external edges and hence there can be
at most two of them also resulting in $T(k-9)$.

If there is at most one extra adjacency, we remove either 19 edges
from which 14 external edges or 20 edges from which 16 external
edges and 7 vertices. Since each tree uses at least three external
edges this results in $T(k-9)$ or better.

\subsubsection{Three degree 4 vertices only two of which are adjacent}
This argument goes in entirely the same way.
Let $x'$, $y'$ and $z'$ be the result of folding $x$, $y$ and $z$ after discarding $v$.
Without loss of generality, assume that $x'$ is adjacent to $y'$ and that $z'$ is not adjacent to any of the other two.
Again we can apply Lemma~\ref{lemdeg3} to the case where we take $v$.
Combined with a $T(k) \leq T_4(k-3) + T(k-9)$ branch or an even better $T(k) \leq T(k-4) + T(k-9)$ branch after discarding $v$,
this leads to a worst case of $T(k) \leq 2T(k-8) + T(k-11) + 2T(k-12)$.

If we discard $x'$, we remove 4 edges and 1 vertex. Now, either a
degree 4 vertex remains giving the $T_4(k-3)$, or an extra edge is
removed by folding giving $T(k-4)$. If we take $x'$, we can also
discard $z'$ resulting in the removal of 17 edges from which 13
external edges and 6 vertices. In the last case there can be at most
one extra adjacency, namely between $z'$ and a degree 3 neighbor of
$x'$. Any tree vertex must again be adjacent to vertices that are
distance at least 3 away from each other in this structure. This can
only be both $z'$ and any neighbor of $x'$. Hence there cannot be
any 1-tree: it would need two neighbors of $x'$ which causes a
4-cycle. Actually there can be no tree at all since every tree leaf
needs to be adjacent to $z'$ in order to avoid 4-cycles in $N(x')$,
but this also implies a 4-cycle. Hence we have
$T(k-17+6+1)=T(k-10)$.

If there is no extra adjacency, there can again be no 1-tree since
it can be adjacent to at most one neighbor of $x'$. Larger trees
remove enough external edges to prove $T(k-9)$.

\subsubsection{Three degree 4 vertices on a path}

Again, we can apply Lemma~\ref{lemdeg3} to the case where we take $v$ which, combined with a $T(k) \leq T_4(k-3) + T(k-9)$ or better branch after discarding $v$,
leads to $T(k) \leq 2T(k-8) + T(k-11) + 2T(k-12)$.
Let $x'$, $y'$ and $z'$ be the result of folding $x$, $y$ and $z$ after discarding $v$,
let $y'$ be adjacent to both $x'$ and $z'$, and let $x'$ and $z'$ be non-adjacent.

If we discard $x'$, we remove 4 edges and 1 vertex while $z'$
remains of degree 4 giving the $T_4(k-3)$. If we take $x'$, we can
also discard $z'$ resulting in the removal of 16 edges from which 11
external edges and 6 vertices. Notice that in the last branch there
cannot be any extra adjacencies since they imply triangles or
4-cycles. There cannot be any trees consisting of 1 or 2 vertices
also because tree leaves can only be adjacent to $z'$ and a degree 3
neighbor of $x'$. Any larger tree decreases the number of external
edges enough to obtain $T(k-16+6+1)=T(k-9)$.

\subsubsection{Folding results in two degree 3 vertices and a non-adjacent a degree 4 vertex}
We now have a graph of complexity $k-3$ with two degree 3 vertices
$y'$, $z'$ and a degree 4 vertex $x'$ which are all the result of
folding. Furthermore, $y'$ and $z'$ are adjacent but not adjacent to
$x'$. Of these vertices $x'$ cannot be involved in any triangle or
4-cycle, or we apply Lemma's~1 case 3 as discussed with the general
approach. Different from before, vertices $y'$ and $z'$ can be
involved in these local structures.

We branch on $x'$. This leads to $T(k-3-3)$ when discarding $x'$.
Similar to the above cases, we can still discard both $y'$ and $z'$
when taking $x'$ in the independent set. Therefore, taking $x'$
leads to removing 17 edges from which 12 external edges and 7
vertices. If there is an extra adjacency, this is between $y'$ or
$z'$ and a neighbor of $x'$. In this case, there can be at most one
tree since $y'$ and $z'$ together have only 3 external edges left
and every tree leaf can be adjacent to at most one neighbor of $x'$
or a 4-cycle with $x'$ would exist. This leads to
$T(k-3-17+7+1+1)=T(k-11)$. If there is no extra adjacency, every
tree leaf can still be adjacent to no more than one neighbor of
$x'$, which together with the 4 external edges of $y'$ and $z'$ lead
to at most 2 trees and $T(k-11)$.

Together with the $T(k-5)$ branch for taking $v$, this leads to $T(k) \leq T(k-5) + T(k-6) + T(k-11)$, which is good enough.

\subsubsection{Folding results in two degree 3 vertices adjacent to a degree 4 vertex}
We again have a graph of complexity $k-3$ with two degree 3 vertices
$y'$, $z'$ and a degree 4 vertex $x'$ which are all the result of
folding. Furthermore, $y'$ is adjacent to $x'$ and $z'$ while $x'$
and $z'$ are non-adjacent. Of these vertices, $x'$ cannot be
involved in any triangle or 4-cycle since we then apply Lemma's~1 case
3 as discussed with the general approach.

Similar to the previous case, we branch on $x'$ giving $T(k-3-3)$ when discarding $x'$, and we allow $y'$ and $z'$ to be discarded when taking $x'$.
This leads to the removal of 14 edges and 6 vertices in the second branch and we have $T(k) \leq T(k-5) + T(k-6) + T(k-11)$ as before unless there are trees.

If there are trees, observe that every tree leaf can again be
adjacent to at most one neighbor of $x'$, and hence all tree leaves
must be adjacent to $z'$. Also observe that the third neighbor of
$y'$ cannot be adjacent to $x'$ or any of its neighbors. Since $z'$
has only two external edges, this means the only tree that can exist
is a 2-tree with both leaves connected to $z'$ and a different
neighbor of $x'$ not equal to $y'$ (or $z$ dominates a tree vertex).
Notice that this implies a triangle involving the tree and $z'$. In
this case we branch on $y'$. When taking $y'$, we remove 10 edges
and 4 vertices: $T(k-6)$. And when discarding $y'$, the tree forms a
triangle in which by dominance $z'$ is taken in the independent set.
Since we can take at most one of the folded vertices, this also
results in $x'$ being discarded. In total, this results in the
removal of 11 edges and 4 vertices, and in this very specific
structure no trees can exist: $T(k-6)$.

\subsubsection{Three degree 4 vertices that form a clique}
The fact that we can take at most one vertex from $x'$, $y'$ and $z'$ is superfluous information here since they already form a clique.
Also, as we discussed with the general case, we cannot use Lemma~\ref{lemdeg3} after taking $v$ in the independent set.
Hence we cannot apply anything from the general approach here and this looks like a very hard case.
However, this case is easy when observing the following.

Let $v$, $x$, $y$, $z$ and $a,\ldots,f$ be as before. Let without
loss of generality $b$ be adjacent to $c$, $d$ be adjacent to $e$,
$f$ be adjacent to $a$, and let non of the vertices in
$\{a,\ldots,f\}$ be adjacent to each other. Notice that when we
discard $v$ this leads to the required adjacencies and triangle of
degree 4 vertices. This is caused by the fact that
$G[N[v]\cup\{a,\ldots,f\}]$ consists of three 5-cycles that overlap
on $v$ and 6 external edges.

If there is a vertex $u \in V$ with a different local structure than just described, we branch on this vertex and are done.
And, if for every vertex $u \in V$ this local structure exists, then $G$ must equal the dodecahedron which has 20 vertices and can be removed in constant time. The proof of Lemma~\ref{lem3} is now completed.

\subsection{Putting it all together}

Lemma~\ref{lemdeg3} described branching on non-3-regular graphs, Lemma~\ref{lem2} described branching on 3-regular graphs that contain triangles or
4-cycles, and Lemma~\ref{lem3} described branching on other 3-regular graphs.
Considering all these branchings we have $T(k) \leq
T(k-8)+2T(k-10)+T(k-12)+2T(k-14)$ in the worst case. This recurrence
relation is formed by combining Lemmata~1 and~3 and leads to a
running time of $O^*(1.17802^k)$. On average degree 3 graphs this is
$O^*(1.17802^{n/2})=O^*(1.08537^n)$.

\begin{theo}\label{thdeg3}
    \is can be solved in $O^*(1.08537^n)$ in connected graphs of
    average degree at most 3.
\end{theo}

\section{Graphs of average degree at most 4}\label{secdeg4}

We deal in this section with (connected) graph of average degree at
most 4. When $m\leq 3n/2$, then we can solve the problem with our
previous algorithm in time $O^*(\gamma^n)$, where
$\gamma=\runningtime$. If $m>3n/2$, then we can branch on a vertex
of degree at least 4. Then the principle of the algorithm is simple:
we branch on vertices of degree at least four as long as $m>3n/2$,
and then we use the algorithm in $O^*(\gamma^n)$ in the remaining
graph.

In our analysis, we seek an algorithm of complexity
$O^*(\gamma^ny^{m-3n/2})$, with $y$ as small as possible. Of course,
we can use the previous study (in Lemma~\ref{lemdeg3}) on branching
of vertices of degree at least 4, but we can do much better, thanks
to our complexity measure. Indeed, we will see that while branching
on a vertex of degree at least 4:
\begin{itemize}
\item either $m$ decreases a lot (respect to $n$) and the branching
is good, \item or we are able to remove a lot of vertices and edges
while branching; this is also good since, intuitively, we will have
a graph with very few vertices when reaching the case $m\leq 3n/2$.
Applying the $O^*(\gamma^n)$ will be `very' fast. \end{itemize} The
result is formally described and proved in the following
proposition.

\begin{prop} \label{dec1} Assume that an algorithm computes a solution
to \is on graphs of average degree $3$, with running time
$O^*(\gamma^n)$. Then, it is possible to compute a solution to \is
on any graph with running time $O^*(\gamma^nf(\gamma)^{m-3n/2})$,
where $f(\gamma)$ is defined by the largest value $y$ verifying a
set of appropriate inequalities. In particular,
$f(\runningtime)=1.13641$.
\end{prop}

\begin{coro}It is possible to compute a solution to \is on graphs with maximum (or even average) degree is $4$ with running time $O^*(1.1571^n)$
\end{coro}

\pr We prove Proposition~\ref{dec1} by a recurrence on $n$ and $m$.
We seek a complexity of the form $O^*(\gamma^ny^{m-3n/2})$. We know
that when $m=3n/2$ (or equivalently when the graph is 3-regular,
since vertices of degree less than 2 have been eliminated by the
preprocessing), we can solve the problem in $O^*(\gamma^n)$. Now, we
assume that our graph has $m>3n/2$ edges. In particular, there is a
vertex of degree at least 4.

Assume that we perform a branching that reduces the graph by either
$\nu_1$ vertices and $\mu_1$ edges, or by $\nu_2$ vertices and
$\mu_2$ edges. Then our complexity formula is valid for $y$ being
the largest root of the following equality:
\begin{equation}\nonumber
\gamma^ny^{m-3n/2}=\gamma^{n-\nu_1}y^{m-3n/2-\mu_1+3\nu_1/2}+\gamma^{n-\nu_2}y^{m-3n/2-\mu_2+3\nu_2/2}
\end{equation}
or equivalently
\begin{equation}\label{eqB}
1=\gamma^{-\nu_1}y^{-\mu_1+3\nu_1/2}+\gamma^{-\nu_2}y^{-\mu_2+3\nu_2/2}
\end{equation}

Then, when $m>3n/2$, one of the following two situations occurs:
\begin{itemize}
\item Either there is a vertex of degree at least 5: in this case we
reduce the graph either by $\nu_1=1$ vertex and $\mu_1=5$ edges, or
by $\nu_2=6$ vertices and $\mu_2\geq 13$ edges, leading to
$y=1.1226$ (or $\nu_1=4$, $\mu_1=9$, $\nu_2=2$, $\mu_2=8$, which is
even better), see Section~\ref{subsecvertex5};
\item Or the maximum degree is 4: Lemma~\ref{lemdeg3} gives a set of possible reductions
that can be plugged into Equation~(\ref{eqB}). As said before, we
can do much better now, thanks to our complexity measure, using the
fact that, informally, removing a lot of vertices might be also
good.
\end{itemize}

In the following, we consider that the graph has maximum degree 4,
and we denote $u_1$, $u_2$ $u_3$ and $u_4$ the four neighbors of
some vertex $v$. We call inner edge an edge between two vertices in
$N(v)$ and outer edge an edge between a vertex in $N(v)$ and a
vertex not in $N[v]$. We study 4 cases, depending on the
configuration of $N(v)$. Here, we consider that no trees are created
while branching. We deal with trees in Section~\ref{sectreedeg4} and
show that it is never problematic.

\medskip

\noindent
{\bf Case 1.} All the neighbors of $v$ have degree 4.

This case is easy. Indeed, if there are at least 13 edges incident
to vertices in $N(v)$, by branching on $v$ we get $\nu_1=1$,
$\mu_1=4$, $\nu_2=5$ and $\mu_2\geq 13$. This gives $y=1.1358$.

But there is only one possibility with no domination and only 12
edges incident to vertices in $N(v)$: when $u_1,u_2,u_3,u_4$ is a
4-cycle. This case reduces thanks to the following lemma.

\begin{figure}[thb]
\begin{center}
\includegraphics[scale=1]{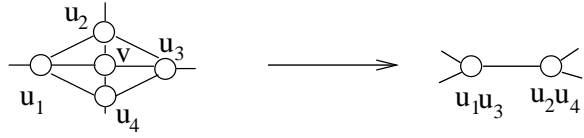}
\caption{$G[N(v)]$ is a 4-cycle}\label{fig1}
\end{center}
\end{figure}

\begin{lem} \label{4cyclesre} Assume there exists a vertex $v$ such that the subgraph
induced by $N(v)$ is a cycle $u_1,u_2,u_3,u_4$. Then, it is possible
to replace $N(v)\cup\{v\}$ by only two vertices $u_1u_3$ and
$u_2u_4$, such that $u$ is adjacent to $u_1u_3$ (resp. $u_2u_4$) if
and only if $u$ is adjacent to $u_1$ or $u_3$ (resp. $u_2$ or
$u_4$).
\end{lem}
\pr Any optimal solution cannot contain more than two vertices from
the cycle. If it contains only one, replacing it by $v$ does not
change its size. Finally, there exist only three disjoint
possibilities: keep $u_1$ and $u_3$, keep $u_2$ and $u_4$ or keep
only $v$, see Figure~\ref{fig1}.~\fdm

\medskip

\noindent
{\bf Case 2.} All the neighbors of $v$ have at least 2 outer edges.

If one of them have degree 4, then there are at least 13 edges
removed when taking $v$, and we get again $\nu_1=1$, $\mu_1=4$,
$\nu_2=5$ and $\mu_2\geq 13$.

Otherwise,once $v$ is removed, any $u_i$ now has degree $2$. Note
that when folding a vertex of degree 2, we reduce the graph by 2 vertices and 2 edges
(if the vertex dominates another one, this is even better). Since any 2 vertices $u_i$
cannot be adjacent to each other, that means we can remove $8$ vertices and at
least $8$ edges by folding $u_1,\cdots,u_4$. Indeed, if for instance $u_1$ dominates
its neighbors (its two neighbors being adjacent), we remove 3 vertices and at least 5
edges which is even better. Removing $8$ vertices and at least $8$ edges is very interesting:
it gives $\nu_1=9$, $\mu_1=12$, $\nu_2=5$, $\mu_2=12$, and $y=1.0856$.

\medskip

\noindent
{\bf Case 3.} $u_1$ has degree 3 and only one outer edge.

$u_1$ has one inner edge, say $(u_1,u_2)$. Let $y$ be the third
neighbor of $u_1$. We branch on $y$. Suppose at first that $u_2$ has
degree 3. If we take $y$ we remove 4 vertices and (at least) 8 edges
(there is at most one inner edge in $N(y)$); if we don't take $y$,
then we remove also $v$ and we remove globally $2$ vertices and $7$
edges.

\begin{figure}[htb]
\begin{center}
\includegraphics[scale=1]{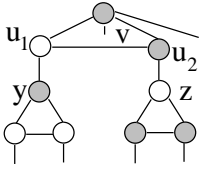}
\caption{Discarding $y$ allows to take $u_1$ and~$z$}\label{fig2}
\end{center}
\end{figure}

This is obviously not sufficient. There is an easily improvable
case, when a neighbor of $y$ has degree 4 (or when $y$ itself has
degree 4), or when the neighbors of $y$ are not adjacent. Indeed, in
this case there are at least 9 edges in $N(y)$, and we get
$\nu_1=4$, $\mu_1=9$, $\nu_2=2$ and $\mu_2\geq 7$, leading to
$y=1.13641$. Now, we can assume that $y$ has degree 3, its 3
neighbors have degree 3. Same for $z$ the neighbor of $u_2$;
furthermore, they both are part of a triangle, see
Figure~\ref{fig2}. Note that $z$ and $y$ cannot be adjacent or there
is a separator of size 2 ($v$ and the third neighbor or $z,y$), and
$z$ and $y$ cannot have a common neighbor (either this vertex would
have degree at least 4, or they have two degree 3 common vertex but
in this case $v$ is a separator 1). At least a neighbor of say $z$
is neither $u_3$ nor $u_4$. Hence, when discarding $y$, we take
$u_1$, so remove $u_2$ and then add $z$ to the solution. Eventually,
we get $\nu_1=4$, $\mu_1=8$, $\nu_2=7$ and $\mu_2\geq 13$, leading
to $y=1.1195$.

Suppose now that $u_2$ has degree 4.  Then, when we don't take $y$,
since we don't take $v$, $u_1$ has degree 1. Then, we can take it
and remove $u_2$ and its incident edges. Then, when not taking $y$,
we remove in all $4$ vertices and $10$ edges. In other words,
$\nu_1=4$, $\mu_1=8$, $\nu_2=4$ and $\mu_2\geq 10$. This gives
$y=1.1325$.

\medskip

\noindent
{\bf Case 4.} $u_1$ has degree 4 and only one outer edge.

Since Case 1 does not occur, we can assume that there is a vertex
(say $u_4$) of degree 3. Since Case 3 does not occur, $u_4$ has no
inner edge. Hence, $u_1$ is adjacent to $u_2$ and $u_3$. Then, there
are only two possibilities.

If there are no other inner edges, since Case 3 does not occur $u_2$
and $u_3$ have 2 outer edges, and we have in all 13 edges. This
gives once again $\nu_1=1$, $\mu_1=4$, $\nu_2=5$ and $\mu_2\geq 13$.

\begin{figure}[htb]
\begin{center}
\includegraphics[scale=1]{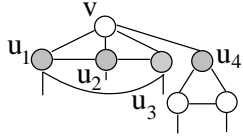}
\caption{$v,u_1,u_2,u_3$ is a 4-clique}\label{fig3}
\end{center}
\end{figure}

Otherwise, there is an edge between $u_2$ and $u_3$. Then,
$v,u_1,u_2,u_3$ form a 4-clique, see Figure~\ref{fig3}. We branch on
$u_4$. If we take $u_4$, we delete $\nu_1=4$ vertices and (at least)
$\nu_2=9$ edges ($v$ has degree 4 and is not adjacent to other
neighbors of $u_4$). If we discard $u_4$, then by domination we take
$v$, and delete $\nu_2=5$ vertices and at least $\mu_2=12$ edges. It
gives $y=1.0921$.

To conclude the proof, we have to verify that removing $\nu'_1$
vertices and $\mu'_1\geq \nu'_1$ edges (without branching) does not
increase the running time. Indeed, this may occur when graph
reductions are performed (such as a vertex folding for instance),
but also in the previous analysis of the possible branchings, since
it may happen that the real reduction remove $\nu_1+\nu'_1$ vertices
and $\mu_1+\mu'_1$ vertices, where $\mu'_1\geq \nu'_1$. To get the
result claimed, we have to verify that $
\gamma^{n-\nu'_1}y^{m-\mu'_1-3n/2+3\nu'_1/2}\leq
\gamma^{n}y^{m-3n/2}$, or equivalently that
$y^{-\mu'_1+3\nu'_1/2}\leq \gamma^{\nu'_1}$. This is trivially true
as soon as $y\leq \gamma^2$ since $\nu'_1\leq \mu'_1$. In other
words, each time we remove $\nu'_1$ vertices and at least $\nu'_1$
edges, we reduce running time with a multiplicative factor
$c^{\nu'_1}$ where $c=(\sqrt{y}/\gamma)<1$.

Similarly, a last issue we have to deal with is what happens if some
branching disconnects our graph. The cases when some trees are
created are handled in Section~\ref{sectreedeg4}. We now assume only
connected components $(C_i)$ each verifying $m_i \geq n_i$ have
appeared. In order to simplify our notation we call $\bar{T}$ the
complexity of
the connected case. 
Our running time now verifies:
\begin{eqnarray*} T(m,n) &\leq& \sum_i \bar{T}(m_i,n_i) =  \sum_i \bar{T}(m-\sum_{j \neq i}m_j,n-\sum_{j \neq i}n_j)\\
&\leq& \sum_i \bar{T}(m-\sum_{j \neq i}(m_j-n_j),n) c^{n-n_i}\leq
\bar{T}(m,n) c^n \sum_i \frac{1}{c^{n_i}}
\end{eqnarray*}
Since $c<1$ (and $n_1\geq 1$), for $n$ large enough we have $c^n
\sum_i \frac{1}{c^{n_i}}\leq 1$ and eventually $T(m,n)\leq
\bar{T}(m,n)$.~\fdm

\section{Dealing with trees}\label{sectreedeg4}

We show here that creating a tree while branching is never
problematic. If we branch on a vertex of degree 3 (as in Case 3),
then no trees are created, or the graph can be reduced without
branching (see Section~\ref{sectreedeg3}).

Now, we consider the case where one or several tree(s) is/are
created when branching on a vertex of degree 4. We denote
$\mathcal{N}$ the number of edges incident to some vertex in $N(v)$,
 $I$ (resp. $\Omega$) the set of edges, called inner edges, that have both endpoints in
 $N(v)$ (resp. the set of edges that have only one endpoint in $N(v)$).

At first, suppose that one of the $l\geq 1$ trees created is a
single vertex $t$. Then, $t$ is a mirror of $v$: when discarding
$v$, we can also discard $t$ and this removes 2 vertices and (at
least) 7 edges. It is easy to see that no tree are created: indeed,
this does not disconnect the graph since at least 3 $u_i$'s are
connected to the remainder of the graph ({i.e.}, the graph after
removing $v$, $N(v)$ and the trees), each tree is connected to at
least one of these 3 $u_i$'s, and the fourth $u_i$ has to be
connected either to 2 trees or to some $u_i$ (or to the remainder of
the graph). When taking $v$, we remove 5 vertices and
$\mathcal{N}\geq |\Omega|\geq 4+3+3l$ edges (since there are at
least 3 edges per tree, and 3 edges to the remainder of the graph).
The reduction of the trees allow to delete at worst $l$ more
vertices (actually $l+k$ vertices and $k$ edges, for some $k\geq
0$). In all, we have $\nu_1=2$, $\mu_1=7$, $\nu_2=5+l$ and
$\mu_2=7+3l$, which is good for $l\geq 2$ (it gives $y=1.1078$). If
$l=1$, then $\mathcal{N}\geq |\Omega|+\lceil
(12-|\Omega|)/2\rceil\geq 10+1=11$, hence the reduction we get is
$\nu_1=2$, $\mu_1=7$, $\nu_2=6$ and $\mu_2=11$. It gives $y=1.1299$.

Now, if $l\geq 2$ and each tree has at least 2 vertices, there are
at least 4 edges linking each tree to $N(v)$. When taking $v$, we
remove 5 vertices and at least $4+3+4l$ edges. When reducing the
trees, we remove additional $2l$ vertices and $l$ edges. In all, we
remove $5+2l$ vertices and $7+5l$ edges. This is of course worse for
$l=2$, for which we have $\nu_1=1$, $\mu_1=4$, $\nu_2=9$ and
$\mu_2=19$ (and $y=1.1031$).

Now, consider the final case where one tree $T$ composed by at least
2 vertices is created while branching on $v$. Then, we have at least
4 edges linking $N(v)$ to $T$, and 3 edges linking $N(v)$ to the
remainder of the graph. Then, $\mathcal{N}=|\Omega|+|I|\geq 11+|I|$.
When taking $v$, since we reduce a tree $T$ of at least 2 vertices,
we delete at worse 7 vertices and $\mathcal{N}+1$ edges.
\begin{itemize}
    \item If all neighbors of $v$ have degree 4, then $\mathcal{N}\geq
    11+\lceil (16-11)/2\rceil=14$. In this case, $\nu_1=1$, $\mu_1=4$, $\nu_2=7$ and $\mu_2=15$.
    It gives $y=1.1315$.
    \item If 1 neighbor of $v$ have degree 3 (and 3 have degree 4), then if there exists at most one inner edge, then $|\Omega|\geq 13$ and
    $\mathcal{N}\geq 14$. Hence, we get at worse $\nu_1=1$, $\mu_1=4$, $\nu_2=7$ and $\mu_2=15$.
    Now, suppose there are two inner edges (hence the tree has two degree 3 vertices $t_1,t_2$). If a vertex $t_1$ of the
    tree is a mirror of $v$, then when discarding $v$ we can discard
    $t_1$ also and get $\nu_1=2$, $\mu_1=7$ (this does not create
    tree). With $\nu_2=7$ and $\mu_2=14$, it gives $y=1.0952$.
    Now, there are only three possibilities without mirror. The first two possibilities occur when the
    two inner edges are $(u_1,u_2)$ and $(u_3,u_4)$. If say $u_3$ is adjacent to
    both $t_1$ and $t_2$ (then $t_1$ is adjacent to $u_1$ and $t_2$ to $u_2$), it is never
    interesting to take $u_3$ (we cannot take 3 vertices if we take $u_3$). The case where $t_1$ is
    adjacent to $(u_1,u_3)$ and $t_2$ to $(u_2,u_4)$
    reduces as follows: we can replace the whole subgraph by two
    adjacent vertices $u_1u_3$ and $u_2u_4$ since either we
    take two vertices $v$ and $t_1$, or we take 3 vertices
    $u_1,u_3,t_2$, or $u_2,u_4,t_1$.
    If the inner edges are $(u_1,u_2)$ and $(u_2,u_3)$, then to
    avoid mirror $u_2$ must be adjacent to say $t_1$, and then $t_1$
    has to be adjacent to $u_4$, and $t_2$ to $u_1$ and $u_3$. But,
    as previously, this case reduces by replacing the whole graph by two
    adjacent vertices $u_1u_3$ and $u_2u_4$.
    \item If 2 neighbors of $v$ have degree 4, and 2 have degree
    3, then there exists at most one inner edge. If there is no inner edge, then $\mathcal{N}=|\Omega|=14$
    and $\nu_1=1$, $\mu_1=4$, $\nu_2=7$ and $\mu_2=15$. If there is
    one inner edge $(u_1,u_2)$, then if $u_3$ or $u_4$ has degree 3,
    when discarding $v$ we can fold two (non adjacent) vertices or
    degree 2. This gives $\nu_1=5$, $\mu_1=8$, $\nu_2=7$ and
    $\mu_2=14$ ($y=1.1244$). If $u_1$ and $u_2$ have degree 3, then to avoid
    separators of size 2 $u_1$ is adjacent to say $t_1$ and $u_2$ is
    not adjacent to the tree. Then, $t_2$ is a mirror of $v$ and we
    get a reduction $\nu_1=2$, $\mu_1=7$, $\nu_2=7$ and $\mu_2=14$.
    \item If one neighbor has degree 4 and the other neighbors of $v$
    have degree 3, then there cannot exist more than one inner edge (because of the degrees).
    If there is no inner edge, then $\mathcal{N}\geq 13$ and,
    as previously, by folding say the 3 (pairwise non adjacent) vertices of degree 3 when not taking $v$, we get
    $\nu_1=7$, $\mu_1=10$, $\nu_2=7$ and $\mu_2=14$ ($y=1.0946$).
    If there is one inner edge $(u_1,u_2)$, then we do not need to branch.
    Indeed, the tree has only two vertices $t_1,t_2$ of degree 3 (otherwise there would be 12 edges in $\Omega$). If say $t_1$
    is adjacent to both $u_1$ and $u_2$, to avoid domination $u_1$ and $u_2$ have to be adjacent to a fourth edge.
    If $t_1$ is adjacent to both $u_3$ and $u_4$, then it is never interesting to
    take $t_1$: indeed, it is impossible to take $t_1$ plus 2 other vertices, and
    we can always take $v$ and $t_2$. If $t_1$ is adjacent to $u_1$ and $u_3$ and $t_2$ to $u_2$ and $u_3$, then at least 2 vertices
    among $u_1,u_2,u_3$ have degree 4 since 2 of them must be adjacent to the remainder of the graph.
    The only remaining case occurs when $t_1$ is adjacent to $u_1,u_3$
    and $t_2$ is adjacent to $u_2,u_4$. In this case we can replace
    the whole subgraph by two adjacent vertices $u_1u_3$ and
    $u_ 2u_4$. Indeed, either we take 2 vertices ($v$ and $t_1$), or
    we take 3 vertices (either $u_1,u_3,t_2$, or $u_2,u_4,t_1$).
    \item Eventually, if all neighbors of $v$ have degree 3, since $|\Omega|\geq 11$, we have $|I|=0$, hence
    $\mathcal{N}=|\Omega|=12$.
    In this case, when we do not take $v$, we have 4 vertices of degree 3
    pairwise non adjacent. We can fold each of them (if there is a
    domination this is even better) and delete 8 more vertices and
    edges. Finally, we get at worse $\nu_1=9$, $\mu_1=12$, $\nu_2=7$ and $\mu_2=14$ ($y=1.0386$). 
\end{itemize}

\section{Graphs of average degree at most 5}\label{secdeg5}

We now consider graphs of average degree 5. We use as in the
previous section a complexity measure that is parameterized by the
algorithm on average degree 4.

More precisely, we proceed as follows. We first identify in
Lemma~\ref{epdeg5} a property linking the average degree of the
graph to the quality of the branching that is performed. Informally,
the bigger the average degree, the more deleted edges when branching
on a (well chosen) vertex. With this property, we analyze the
complexity of our algorithm in a bottom up way as follows. If we
know how to solve the problem in $O^*(\gamma^n)$ in graph with
average degree $d$, and that when the average degree is greater than
$d$ a good branching occurs, we seek a complexity of the form
$O^*(\gamma^ny^{m-dn/2})$, valid in graph with average degree
greater than $d$. Starting from $d=4$, we identify four critical
values for the average degree, leading to a complexity of
$O^*(1.1969^n)$ in graphs of average degree at most 5.

\begin{lem} \label{epdeg5} Assume the input graph has maximum degree $5$ and average degree $4$ or more. Then
$$T(m,n)\leq T(n-1,m-5)+T(n-6,m-15)$$
Or some even better case happens. Furthermore, if it verifies:
\begin{itemize}
    \item  $m > 20n/9$, then $T(n,m)\leq T(n-1,m-6)+T(n-7,m-16)$
    \item  $m > 16n/7$, then $T(n,m)\leq T(n-1,m-6)+T(n-7,m-17)$
    \item  $m > 12n/5$, then $T(n,m)\leq T(n-1,m-6)+T(n-7,m-18)$
\end{itemize}

\end{lem}

\pr Fix some vertex $v_0$ of degree $5$, such that for any vertex
$v$ of degree $5$ in the graph:
\begin{equation*}
\sum_{w \in N(v)}d(w) \leq  \sum_{w \in N(v_0)}d(w) = \delta
\end{equation*}
For $i\leq 5$, let $m_{i5}$ be the number of edges in the graph
between a vertex of degree $i$ and a vertex of degree 5. For $i\leq
4$, fix $\alpha_i = m_{i5}/n_5$ and $\alpha_5 = 2m_{55}/n_5$. In
other terms, $\alpha_i$ is the average number of vertices of degree
$i$ that are adjacent to a vertex of degree $5$. However, we can
always consider $\alpha_i=0$ for $i\leq 2$. Summing up inequalities
on any vertex of degree $5$, we get:
\begin{eqnarray}
\label{alpha01} \sum_{i\leq 5}i\alpha_i &\leq& \delta\\
\label{alpha02} \sum_{i\leq 5}\alpha_i &=& 5
\end{eqnarray}
Fix now $\epsilon = m/n-2 \in ]0,1/2[$.
\begin{equation*}
\epsilon = \frac{n_5-n_3}{2(n_5+n_4+n_3)}
\end{equation*}
This function is decreasing with $n_3$ and $n_4$. We now use some
straightforward properties:
\begin{eqnarray*}
n_4 &\geq& \frac{m_{45}}{4}\\
n_3 &\geq& \frac{m_{35}}{3}\\
5n_5 &=& m_{35}+m_{45}+2m_{55}
\end{eqnarray*}
That leads us to:
\begin{eqnarray*}
\epsilon &\leq& \frac{3n_5-m_{35}}{6n_5+\frac{3}{2}m_{45}+2m_{35}}\\
&\leq& \frac{m_{45}+2m_{55}-2n_5}{16n_5-\frac{1}{2}m_{45}-4m_{55}}
\end{eqnarray*}
And, by hypothesis:
\begin{equation} \label{epsilmax01}
\epsilon \leq \frac{2\alpha_4+2\alpha_5-4}{32-\alpha_4-4\alpha_5}
\end{equation}

Let $\mu_2$ be the minimal number of edges we delete when we add
$v_0$ to the solution. Since there are at least $2d(v_0)$ edges
between $N(v_0)$ and the remaining of the graph, and thanks to
inequalities (\ref{alpha01}) and (\ref{alpha02}), we get:
\begin{eqnarray*}
\mu_2 &\geq& 10+\left\lceil\frac{\delta-10}{2}\right\rceil\\\
&\geq& 10+\left\lceil\frac{5+\alpha_4+2\alpha_5}{2}\right\rceil
\end{eqnarray*}

Notice that $\epsilon >0$ implies:
\begin{equation}\label{epsilonpos01}\alpha_4+\alpha_5 > 2\end{equation}
If we run $\min \mu_2$ under constraints
(\ref{alpha01}),(\ref{alpha02}),(\ref{epsilonpos01}) and $\mu_2 \in
\mathbb{N}$, we find $\mu_2=14$ as a minimum.

For $1\leq i\leq 3$, we now consider the following programs $(P_i)$:
$\max \epsilon$ under constraints
(\ref{alpha01}),(\ref{alpha02}),(\ref{epsilmax01}) and $\mu_2 \leq
14+i$. In other terms, we determine the maximal value for $\epsilon$
such that it is possible that no vertex in the graph verifies
$\mu_2=15+i$. The following table summarizes the results:
$$
\begin{array}{c|c|c}
\text{worst case for }\mu_2 & \text{upper bound for }\epsilon & (\alpha_5,\alpha_4)\\
\hline \hline
14 & 2/29 & (0,3)\\
15 & 2/9 & (0,5)\\
16 & 2/7 & (2,3)\\
17 & 2/5 & (4,1)
\end{array}
$$

Notice also that $\mu_2=14$ implies that at least one neighbor of
$v_0$ has degree 3, so we can fold it after discarding $v$. In that
case, we get $\nu_1=3,\mu_1=7,\nu_2=6,\mu_2=14$, that is better than
$\nu_1=1,\mu_1=5,\nu_2=6,\mu_2=15$.\fdm

\begin{prop} \label{findeg5} Assume that an algorithm computes a solution to \is on graphs with average degree at most $4$, with running time $O^*(\gamma^n_0)$. Then, it is possible to compute a solution to \is on any graph with running time:
$$O^*(\gamma^n_0 \gamma_1^{2n/9} \gamma_2^{4n/63} \gamma_3^{4n/35} \gamma_4^{m-2n/5})$$ for some appropriate constants $(\gamma_i)_{i \leq 4}$. 
In particular,
$$
\gamma_0=1.1571 \Longrightarrow \left\{\begin{array}{lcl}
\gamma_1 = 1.0775 \\
\gamma_2 = 1.0696 \\
\gamma_3 = 1.0631 \\
\gamma_4 = 1.0612
\end{array}\right.
$$
\end{prop}

To be more precise, $\gamma_i$ corresponds to the case where our
graph is dense enough to state that $\mu_2 \geq 15+i$, according to
Lemma~\ref{epdeg5} (the case when there is a vertex of degree at
least 6 can be easily shown to lead to a better reduction).

\medskip

\pr We seek a complexity of the form
$O^*(\gamma^ny^{m-(2+\epsilon_i) n})$, where $2+\epsilon_i$ is the
lowest ratio $m/n$ that allows us to use $\nu_1=1$, $\mu_1=6$,
$\nu_2=7$ and $\mu_2=14+i$ in the recurrence equation:
\begin{equation}\label{eqBG}
1=\gamma^{-\nu_1}y^{-\mu_1+(2+\epsilon_i)\nu_1}+\gamma^{-\nu_2}y^{-\mu_2+(2+\epsilon_i)\nu_2}
\end{equation}
According to Lemma~\ref{epdeg5}, we know that
$$(\epsilon_i)_{i \leq 4} = (0,2/9,2/7,2/5).$$ In the worst
case this leads to the values summarized in the table just above.

Note that a reduction of $\nu'_1$ vertices and $\mu'_1\geq \nu'_1$
edges is not problematic for $y\leq \gamma_i^{1/(1+\epsilon_i)}$.

In order to deal with trees, note also that removing a tree
corresponds to a reduction of $\nu$ vertices and $\nu-1$ edges. This
is not problematic as soon as $y^{1.5\nu+1}\leq \gamma^{\nu}$. This
is true for $\nu\geq 2$.

Otherwise, trees are singletons and there is no separator of size
$2$ or less. We also get $|\Omega| \geq 5+3+3l$ , that means $\mu_2
\geq 8+2l+\left\lceil\frac{\delta-8-3l}{2}\right\rceil$. Hence, we
see that if $l>1$, or if there are at least 4 edges linking vertices
in $N(v)$ to the remainder of the graph, or if our disconnected
vertex $t$ has degree at least $4$, we are in a better situation as
when no tree is created. Eventually, assume $d(t)=3$ and there is a
separator of size $3$, namely $u_3$, $u_4$ and $u_5$. $t$ is
adjacent to $u_1$,$u_2$  and, say, $u_3$. If $u_1$ and $u_2$ are not
adjacent, then it is never interesting to take $v$ (if we take $v$
we take only $t$ in $N(v)\cup\{v,t\}$, and we can take $u_1,u_2$
instead). Otherwise, no more than $3$ vertices from $N(v)$ may
belong to the optimal (otherwise that would mean for instance
$N(v)-u_1$ contains no edge, and thus $u_1$ dominates $u_2$), and
there are only $3$ different ways to choose $2$ vertices among
$u_3,u_4,u_5$. So we can replace the whole subgraph by a clique of
size at most $3$.~\fdm

\begin{theo} It is possible to compute a solution to \is on graph whose maximum (or even average) degree is $5$ with running time $O^*(1.1969^n)$
\end{theo}

\pr We just apply Proposition~\ref{findeg5} with $m \leq 5n/2$~\fdm

\section{Graphs of average degree at most 6}\label{secdeg6}

We apply here a technique similar to the case of graphs with average
degree at most 6.

\begin{lem} \label{epdeg6} Assume the input graph has maximum degree $6$ and average degree $5$ or more. Then
$$T(m,n)\leq T(n-1,m-6)+T(n-7,m-20)$$
Furthermore, if it verifies:
\begin{itemize}
    \item  $m > 60n/23$, then $T(n,m)\leq T(n-1,m-6)+T(n-7,m-21)$
    \item  $m > 60n/22$, then $T(n,m)\leq T(n-1,m-6)+T(n-7,m-22)$
    \item  $m > 205n/74$, then $T(n,m)\leq T(n-1,m-6)+T(n-7,m-23)$
    \item  $m > 20n/7$, then $T(n,m)\leq T(n-1,m-6)+T(n-7,m-24)$
\end{itemize}

\end{lem}

\pr Fix some vertex $v_0$ of degree $6$, such that for any vertex of
degree $6$ in the graph:
\begin{equation*}
\sum_{w \in N(v)}d(w) \leq  \sum_{w \in N(v_0)}d(w) = \delta
\end{equation*}
For $i\leq 5$, fix $\alpha_i = m_{i6}/n_6$ and $\alpha_6 =
2m_{66}/n_6$. In other terms, $\alpha_i$ is the average number of
vertices of degree $i$ that are adjacent to a vertex of degree $6$.
However, we can always consider $\alpha_i=0$ for $i\leq 2$. Summing
up inequalities on any vertex of degree $6$, we get:
\begin{eqnarray}
\label{alpha1} \sum_{i\leq 6}i\alpha_i &\leq& \delta\\
\label{alpha2} \sum_{i\leq 6}\alpha_i &=& 6
\end{eqnarray}
Fix now $\epsilon = m/n-5/2 \in ]0,1/2[$.
\begin{equation*}
\epsilon = \frac{n_6-n_4-2n_3}{2(n_6+n_5+n_4+n_3)}
\end{equation*}
This function is decreasing with $n_3$,$n_4$ and $n_5$. We now use
some straightforward properties:
\begin{eqnarray*}
n_5 &\geq& \frac{m_{56}}{5}\\
n_4 &\geq& \frac{m_{46}}{4}\\
n_3 &\geq& \frac{m_{36}}{3}\\
6n_6 &=& m_{36}+m_{46}+m_{56}+2m_{66}
\end{eqnarray*}
That leads us to:
\begin{eqnarray*}
\epsilon &\leq& \frac{60n_6-15m_{46}-40m_{36}}{120n_5+24m_{56}+30m_{46}+40m_{36}}\\
&\leq&
\frac{25m_{46}+40m_{56}+80m_{66}-180n_6}{360n_6-10m_{36}-16m_{56}-80m_{66}}
\end{eqnarray*}
And, by hypothesis:
\begin{equation} \label{epsilmax}
\epsilon \leq
\frac{25\alpha_4+40\alpha_5+40\alpha_6-180}{360-10\alpha_4-16\alpha_5-40\alpha_6}
\end{equation}

Once again, let $\mu_2$ be the minimal number of edges we delete
when we add $v_0$ to the solution. Since there are at least
$2d(v_0)$ edges between $N(v_0)$ and the remaining of the graph, and
thanks to inequalities (\ref{alpha1}) and (\ref{alpha2}), we get:
\begin{eqnarray*}
\mu_2 &\geq& 12+\left\lceil\frac{\delta-12}{2}\right\rceil\\\
&\geq&
15+\left\lceil\frac{\alpha_4+2\alpha_5+3\alpha_6}{2}\right\rceil
\end{eqnarray*}

Notice that $\epsilon >0$ implies:
\begin{equation}\label{epsilonpos}5\alpha_4+8\alpha_5+8\alpha_6 > 36\end{equation}
If we run $\min \mu_2$ under constraints
(\ref{alpha1}),(\ref{alpha2}),(\ref{epsilonpos}) and $\mu_2 \in
\mathbb{N}$, we find $\mu_2=20$ as a minimum, that proves our first
claim. (limit case $\mu=19$ and $\epsilon=0$ is reached when
$\alpha_6=0$, $\alpha_5=2$ and $\alpha_4=4$)

For $1\leq i\leq 4$, we now consider the following programs $(P_i)$:
$\max \epsilon$ under constraints
(\ref{alpha1}),(\ref{alpha2}),(\ref{epsilmax}) and $\mu_2 \leq
19+i$. In other terms, we determine the maximal value for $\epsilon$
such that it is possible that no vertex of degree 6 in the graph
verifies $\mu_2=20+i$. The following table summarizes the results
and concludes the proof of the lemma.~\fdm
$$
\begin{array}{c|c|c}
\text{Worst case for }\mu_2 & \text{Upper bound for }\epsilon & (\alpha_6,\alpha_5,\alpha_4)\\
\hline \hline
20 & 5/46 & (0,4,2)\\
21 & 5/22 & (0,6,0)\\
22 & 10/37 & (2,4,0)\\
23 & 5/14 & (4,2,0)
\end{array}
$$
\begin{prop} \label{findeg6} Assume that an algorithm computes a solution to \is on graphs with average degree at most $5$, with running time $O^*(\gamma^n_0)$. Then, it is possible to compute a solution to \is on any graph with running time:
$$O^*(\gamma^n_0 \gamma_1^{5n/46} \gamma_2^{85n/252} \gamma_3^{35n/814} \gamma_4^{45n/518} \gamma_5^{m-5n/14})$$ for some appropriate constants $(\gamma_i)_{i \leq 5}$. 
In particular,
$$
\gamma_0=1.1969 \Longrightarrow \left\{\begin{array}{lcl}
\gamma_1 & = & 1.0356 \\
\gamma_2 & = & 1.0327 \\
\gamma_3 & = & 1.0301 \\
\gamma_4 & = & 1.0278 \\
\gamma_5 & = & 1.0258
\end{array}\right.
$$
\end{prop}

To be more precise, $\gamma_i$ corresponds to the case where our
graph is dense enough to state that $\mu_2 \geq 19+i$, according to
Lemma~\ref{epdeg6}.

\medskip

\pr We seek a complexity of the form
$O^*(\gamma^ny^{m-(5/2+\epsilon_i) n})$, where $5/2+\epsilon_i$ is
the lowest ratio $m/n$ that allows us to use $\nu_1=1$, $\mu_1=6$,
$\nu_2=7$ and $\mu_2=19+i$ in the recurrence equation:
\begin{equation}\label{eqBG2}
1=\gamma^{-\nu_1}y^{-\mu_1+(5/2+\epsilon_i)\nu_1}+\gamma^{-\nu_2}y^{-\mu_2+(5/2+\epsilon_i)\nu_2}
\end{equation}
According to Lemma~\ref{epdeg6}, we know that
$$(\epsilon_i)_{i \leq 5} = (0,5/46,5/22,10/37,5/14).$$ In the worst
case this leads to the values summarized in the table just above.
Note that a reduction of $\nu'_1$ vertices and $\mu'_1\geq \nu'_1$
edges is not problematic for $y\leq \gamma_i^{2/(3+2\epsilon_i)}$.

In order to deal with trees, note also that removing a tree
corresponds to a reduction of $\nu$ vertices and $\nu-1$ edges. This
is not problematic as soon as $y^{2.5\nu+1}\leq \gamma^{\nu}$. This
is true for $\nu\geq 1$. In other words, removing a tree reduces the
global complexity.~\fdm

\begin{theo} It is possible to compute a solution to \is on graph whose maximum (or even average) degree is $6$ with running time
$O^*(1.2149^n)$.
\end{theo}

\pr We just apply Proposition~\ref{findeg6} with $m \leq 3n$.~\fdm

\section{Conclusion}

We have tackled in this paper worst-case complexity for \is{} in graphs with average degree~3, 4, 5 and~6. The results obtained improve upon the best results known for these problems. Let us note that the cases of average degrees~5 and~6 deserve further refinement. Indeed, it seems to us that there is enough place for improving them, since our results are got by using fairly simple combinatorial arguments.

An interesting point of our work is that improvement for the three last cases have been derived based upon a new method following which any worst-case complexity result for \is{} in graphs of average degree~$d$ can be used for deriving worst-case complexity bounds in any graph of average degree greater than~$d$. This method works for any average degree's value and can be used for any graph-problem where the larger the degree the better the worst-case time-bound obtained.

\bibliographystyle{plain}

\end{document}